\documentclass[a4paper,11pt]{article}
\pdfoutput=1 

\usepackage{jcappub} 

\usepackage[T1]{fontenc} 
\usepackage{aas_macros}
\usepackage[utf8]{inputenc}
\usepackage{graphicx,color,amssymb,amsmath,mathtools,cases,bm,float,tocbibind,subfig,mathabx, xcolor}
\usepackage{enumitem}
\usepackage[format=hang,font=small,textfont=it]{caption}

\newcommand{\beq}{\begin{equation}}
\newcommand{\eeq}{\end{equation}}
\newcommand{\beqs}{\begin{eqnarray}}
\newcommand{\eeqs}{\end{eqnarray}}

\title{Performance forecasts for the primordial gravitational wave detection pipelines for AliCPT-1}

\author[a,b]{Shamik Ghosh,}
\author[c]{Yang Liu,}
\author[d,e,p]{Le Zhang,}
\author[c]{Siyu Li,}
\author[d]{Junzhou Zhang,}
\author[f]{Jiaxin Wang,}
\author[a,b]{Jiazheng Dou,}
\author[a,b]{Jiming Chen,}
\author[g]{Jacques Delabrouille,}
\author[i]{Mathieu Remazeilles,}
\author[a,b]{Chang Feng,}
\author[j]{Bin Hu,}
\author[d,e]{Zhi-Qi Huang,}
\author[k,c]{Hao Liu,}
\author[l]{Larissa Santos,}
\author[f,m,n]{Pengjie Zhang,}
\author[d]{Zhaoxuan Zhang,}
\author[a,b,1]{Wen Zhao,\note{Corresponding author.}}
\author[c,1]{Hong Li,}
\author[o]{Xinmin Zhang}

\affiliation[a]{CAS Key Laboratory for Researches in Galaxies and Cosmology, Department of Astronomy, University of Science and Technology of China, Chinese Academy of Sciences, Hefei, Anhui 230026, People's Republic of China}
\affiliation[b]{School of Astronomy and Space Science, University of Science and Technology of China, Hefei 230026, People's Republic of China}
\affiliation[c]{Key Laboratory of Particle Astrophysics, Institute of High Energy Physics, Chinese Academy of Sciences, 19B Yuquan Road, Beijing 100049, People’s Republic of China}
\affiliation[d]{School of Physics and Astronomy, Sun Yat-sen University, 2 Daxue Road, Tangjia, Zhuhai, 519082, People's Republic of China}
\affiliation[e]{CSST Science Center for the Guangdong-Hong Kong-Macau Greater Bay Area, Zhuhai 519082, People's Republic of China}
\affiliation[f]{Department of Astronomy, School of Physics and Astronomy, Shanghai Jiao Tong University, Shanghai, 200240, People's Republic of China}
\affiliation[g]{CNRS-UCB International Research Laboratory, Centre Pierre Bin\'etruy, IRL2007, CPB-IN2P3, Berkeley, USA}
\affiliation[i]{Instituto de F\'isica de Cantabria (CSIC-UC), Avenida de los Castros s/n, 39005 Santander, Spain}
\affiliation[j]{Department of Astronomy, Beijing Normal University, Beijing 100875, People's Republic of China}
\affiliation[k]{School of Physics and optoelectronics engineering, Anhui University, 111 Jiulong Road, Hefei, 230601, People's Republic of China}
\affiliation[l]{Center for Gravitation and Cosmology, College of Physical Science and Technology, Yangzhou University, Yangzhou, 225009, People's Republic of China}
\affiliation[m]{Shanghai Key Laboratory for Particle Physics and Cosmology, Shanghai, 200240, People's Republic of China}
\affiliation[n]{Division of Astronomy and Astrophysics, Tsung-Dao Lee Institute, Shanghai Jiao Tong University, Shanghai, 200240, People's Republic of China}
\affiliation[o]{Theoretical devision, Institute of High Energy Physics, Chinese Academy of Sciences, 19B Yuquan Road, Beijing 100049, People’s Republic of China}
\affiliation[p]{Peng Cheng Laboratory, No.2, Xingke 1st Street, Shenzhen 518000, People’s Republic of China}

\emailAdd{wzhao7@ustc.edu.cn}
\emailAdd{hongli@ihep.ac.cn}

\newcommand{\tblue}[1]{\textcolor{blue}{#1}}

\abstract{AliCPT is the first Chinese cosmic microwave background (CMB) experiment which will make the most precise measurements of the CMB polarization in the northern hemisphere. The key science goal for AliCPT is the detection of primordial gravitational waves (PGWs). It is well known that an epoch of cosmic inflation, in the very early universe, can produce PGWs, which leave an imprint on the CMB in form of odd parity $B$-mode polarization. In this work, we study the performance of the component separation and parameter estimation pipelines in context of constraining the value of the tensor-to-scalar ratio. Based on the simulated data for one observation season, we compare five different pipelines with different working principles. Three pipelines perform component separation at map or spectra level before estimating $r$ from the cleaned spectra, while the other two pipelines performs a global fit for both foreground parameters and $r$. We also test different methods to account for the effects of time stream filtering systematics. This work shows that our pipelines provide consistent and robust constraints on the tensor-to-scalar ratio and 
a consistent sensitivity $\sigma(r) \sim 0.02$. This showcases the potential of precise $B$-mode polarization measurement with AliCPT-1. AliCPT will provide a powerful opportunity to detect PGWs, which is complementary with various ground-based CMB experiments in the southern hemisphere.}

\begin{document}
\maketitle
\flushbottom

\section{Introduction to science objectives}
\label{sec:intro}
The $\Lambda$CDM model has been widely successful in explaining the observed universe with a minimum of six independent parameters \cite{Planck2020a}. Despite its success, the $\Lambda$CDM model has some well known issues like the horizon, flatness and the magnetic monopole problems. Inflation is an epoch in the very early universe, when the universe is supposed to undergo a nearly exponential expansion, that alleviates the problems of the $\Lambda$CDM model \citep{Brout1978, Starobinsky1980, Kazanas1980, Sato1981, Guth1981, Linde1982, Linde1983, Steinhardt1982}. It was soon realized that quantum fluctuations during inflation can generate primordial cosmological perturbations \citep{Mukhanov1981, Mukhanov1982, Hawking1982, Guth1982, Starobinsky1982, Bardeen1983, Mukhanov1985}, which would evolve into the large scale structures. Cosmic inflation is currently the leading paradigm for early universe physics.

Measurements of the cosmic microwave background (CMB) radiation have been some of the most important and accurate observations that have helped cosmologists make precise estimations of $\Lambda$CDM model parameters \cite{Planck2020a}. The anisotropies in the CMB intensity are related to the scalar perturbations 
of the spacetime metric
produced in the early universe. The CMB is also linearly polarized. The even parity, $E$-mode polarization of the CMB is produced at last scattering from local quadrupolar anisotropies in the CMB intensity. However, the odd-parity, $B$-mode polarization cannot be produced from scalar perturbations. In the linear order approximation, the $B$-modes can only be produced by primordial gravitational waves (PGWs), i.e. the tensor perturbations of the metric, produced during the epoch of inflation \citep{Hu1997, Kamionkowski1997, Seljak1997,zhao2006,zhao2009,zhao2009b,zhao2009c}. While $B$-modes are also produced by conversion of $E$-modes to $B$-modes through gravitational lensing, or from astrophysical emissions in the foreground, the $B$-modes remain our best bet of finding the so called `smoking-gun' evidence of an inflationary epoch.

We describe the amplitude of the PGW signal in terms of the ratio of the tensor power spectrum to the scalar power spectrum, at the same pivot scale. The current efforts of CMB experiments in observing the PGW signal is focused on making a significant measurement of the tensor-to-scalar ratio $r$. A significant measurement of $r$ would not only serve as a possible observation of a signal from inflation, it can also be used to constrain models of inflation. Currently there is a very large variety in possible models of inflation. These models have widely different predictions for $r$. So a significant measurement of $r$ also helps us narrow down models of inflation.

Over the past decade the CMB community has stepped up efforts to make precise measurements of the CMB $B$-mode polarization. Direct measurement of the lensed $B$-mode signal has been reported by BICEP2 \citep{Bicep2014}, the Keck Array \cite{BK2015}, SPTpol \citep{SPT2015}, POLARBEAR \citep{Polarbear2017}, and ACTpol \citep{ACT2017}. Currently suborbital CMB experiments like ACTpol \citep{ACTpol2020}, BICEP/Keck \citep{BK2022}, CLASS \citep{CLASS2016}, POLARBEAR/Simons Array \citep{SimonsArray2016}, SPTpol \citep{SPTpol2020} are making high sensitivity measurements of the CMB polarization to make a significant detection of the tensor-to-scalar ratio. At present, the tightest limits on the tensor-to-scalar ratio at a pivot scale of $k_* =0.05\text{ Mpc}^{-1}$ is $r<0.036$ at 95\% confidence, which comes from the recent joint analysis of BICEP/Keck and Planck data \citep{BK2022}. Aside from these ongoing experiments, LSPE \citep{LSPE2021}, QUBIC \citep{QUBIC2019}, Simons Observatory \citep{SO2019}, and SPIDER \citep{SPIDER2022} will be joining the measurement efforts soon. The future of CMB $B$-mode measurements will be lead by the LiteBIRD satellite mission \citep{LiteBIRD2019}, and the CMB-S4 ground-based observatory \citep{CMBS42022}.

The Ali CMB Polarization Telescope (AliCPT) is the first Chinese CMB experiment. It will measure the CMB polarization in the northern hemisphere at 90 GHz and 150 GHz, located at Ali, in Tibet, China \citep{AliCPT2017science, AliCPT2020receiver,alicpt-lens}. The key scientific objectives of the AliCPT project include producing the best CMB polarization observations of the northern sky, and making a significant detection of PGW using the high precision polarization data \citep{PaperI}. This work presents a forecast of PGW detection at AliCPT-1. This paper is part of a series of articles aimed at demonstrating the simulation and data analysis pipelines designed to deliver the key science goals of the AliCPT project. In \citep{PaperII} (hereafter Paper II) we detailed the AliCPT-1 simulation pipeline. The foreground cleaning pipelines are discussed in \citep{PaperIII} (hereafter Paper III). This paper builds on Paper II and Paper III to demonstrate the performance of our data analysis methods and make forecast for the tensor-to-scalar ratio, assuming one season of observations.

Presently, data analysis pipelines for PGW constraints mostly adopt a likelihood framework assuming a parametric model for foreground and the CMB \citep{Joint_BICEP&Planck22015, BK2022}. However, recently proposed component separation methods \citep{2019arXiv190807862S, Remazeilles2021} provide alternative approaches to tackle this problem. For AliCPT-1, we have five different pipelines: 1. the analytic blind separation (ABS), 2. the generalized least squares (GLS), 3. the constrained internal linear combination (cILC), 4. the multi-component multi-frequency likelihood (McMfL), and 5. the template fitting (TF) pipelines. These methods approach the problem of detecting the small PGW signal with different principles. {We do not employ either a map-based or a timestream-based likelihood method (like Commander \citep{Eriksen2004b}, ForegroundBuster \citep{Stompor2009}) in our data analysis. While full Bayesian methods of such kind provide an alternate approach in data analysis and error propagation, incorporating timestream filtering effects/corrections in such methods are non-trivial and are an open problem.}

We outline the mock data sets used in our primordial $B$-mode analysis in Sec.~\ref{sec:mock_data}. In Sec.~\ref{sec:Cl_estimators} we briefly outline the power spectrum estimators used in the analysis. A summary of the filtering effects correction is discussed in Sec.~\ref{sec:filter}. In Sec.~\ref{sec:pipelines} we discuss the five pipelines used in our primordial $B$-mode analysis, their parameter estimates and performance. We make a comparative summary of our estimate of the tensor-to-scalar ratio $r$ with AliCPT-1 in Sec.~\ref{sec:conclusions}. 

\section{Mock data sets}
\label{sec:mock_data}
{We summarize the basic information and hardware configuration parameters of the AliCPT experiment in the Table \ref{table:intrument-parameters}. The parameters listed here are for the receiver at full load, i.e., with 19 detector modules, as well as in the first stage, with only 4 modules. Each module has 1704 transition edge sensors (TES). }

\begin{table}[H]
    \caption{Summary table of the basic information of AliCPT of full load and first stage.}
    \label{table:intrument-parameters}
    \centering

\begin{tabular} { | l | c | c | }
\hline
 \textbf{Location of site} & \multicolumn{2}{c|}{$32^{\circ}18^{\prime}38^{\prime\prime}N$, $80^{\circ}1^{\prime}50^{\prime\prime}E$}\\
 \hline
 \textbf{Caliber (mm)} & \multicolumn{2}{c|}{720}\\
  \hline
 \textbf{Field of View ($\circ$)} & \multicolumn{2}{c|}{33.4}\\
 \hline
 \textbf{Detector Technology} & \multicolumn{2}{c|}{TES}\\
 \hline
 \textbf{Sky Coverage} & \multicolumn{2}{c|}{ $60\%$, of which $17\%$ clean patch focus on $r$ }\\
\hline
 \textbf{Channels (GHz)} &  \textbf{90} & \textbf{150}\\ 
\hline
\hline
\textbf{Center Frequency (GHz)} & 91.4 &  145 \\
\textbf{Bandwidth (GHz)} & 38 &  40 \\
\textbf{Resolution (FWHM)} & $19^{\prime}$ & $11^{\prime}$ \\
\textbf{Optical TES count of full load} & 16,188 & 16,188 \\
\textbf{Optical TES count of 1st stage} & 3408 & 3408 \\

\textbf{NET CMB ($\mu K\sqrt{s}$)} & 274 & 348 \\
\hline

\end{tabular}
\end{table}

{The configuration of 4 modules with a total of 6816 detectors is the main configuration of AliCPT in the first phase. After that, the number of detectors will be gradually upgraded. In this paper, simulations will be carried out based on the first phase with 4 modules, and 1 full observation season measurements, and preliminary study results will be given.} We built a dedicated data simulation pipeline to predict the scientific capability of AliCPT experiment. The simulations used in the forecast of primordial gravitational waves were produced with this dedicated simulation pipeline. In this section, we provide a general description of the pipeline processes, and more details can be found in Paper II. 

\begin{figure}
    \centering
    \includegraphics[width=0.48\textwidth, trim= 3.5cm 0cm 3.5cm 0cm, clip=true]{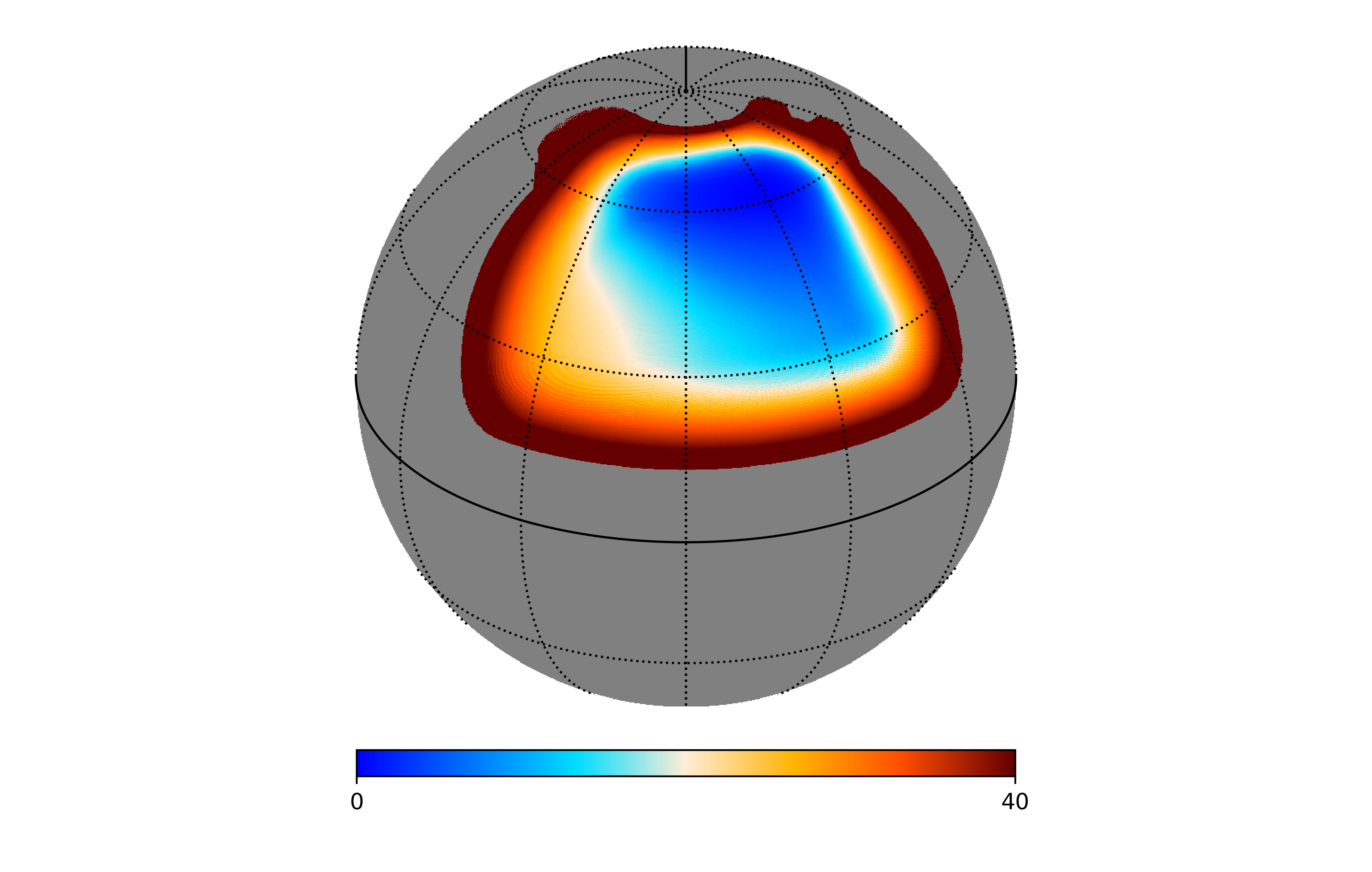}
    \includegraphics[width=0.48\textwidth, trim= 3.5cm 0cm 3.5cm 0cm, clip=true]{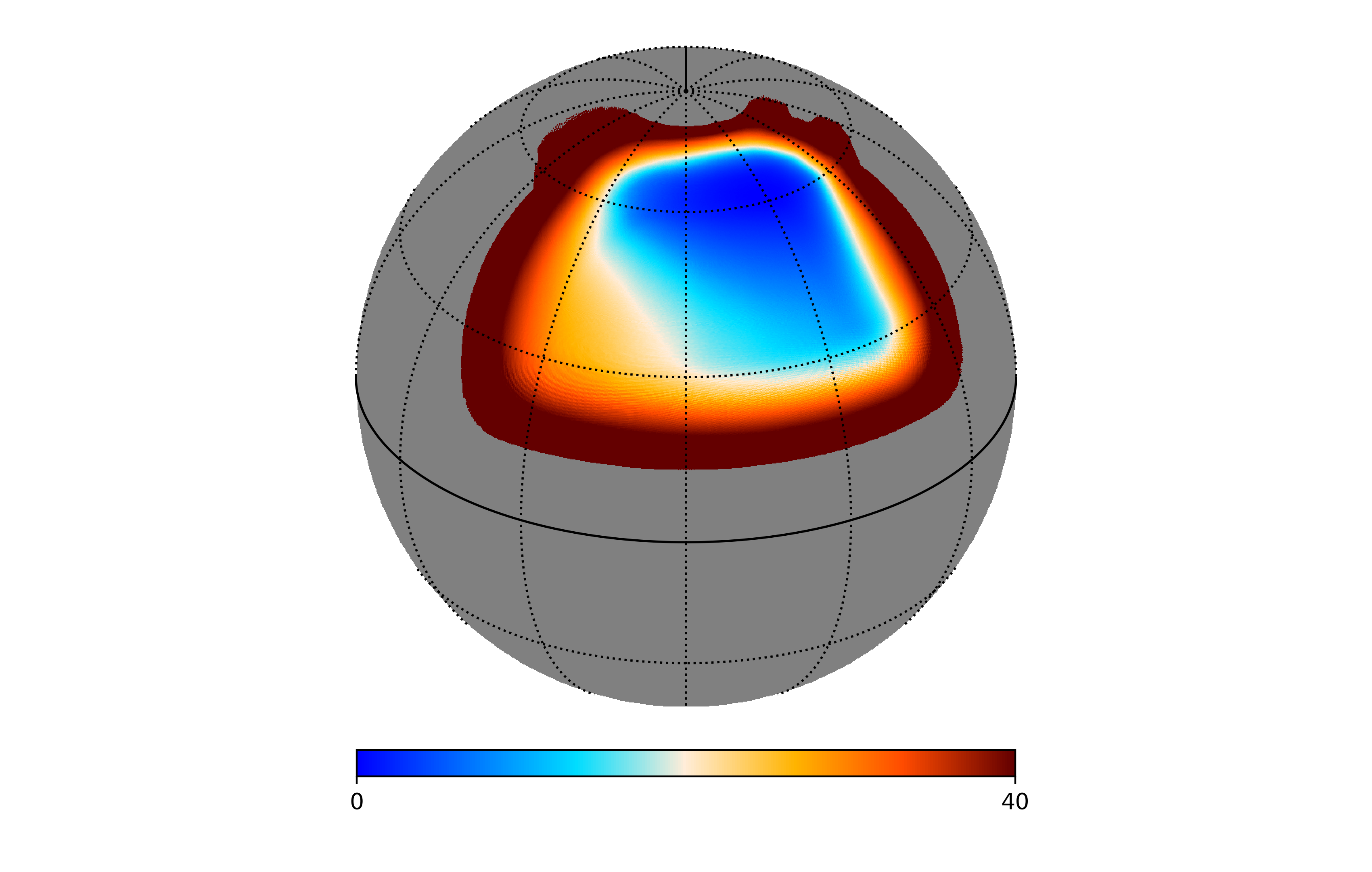}\\
    \includegraphics[width=0.60\textwidth]{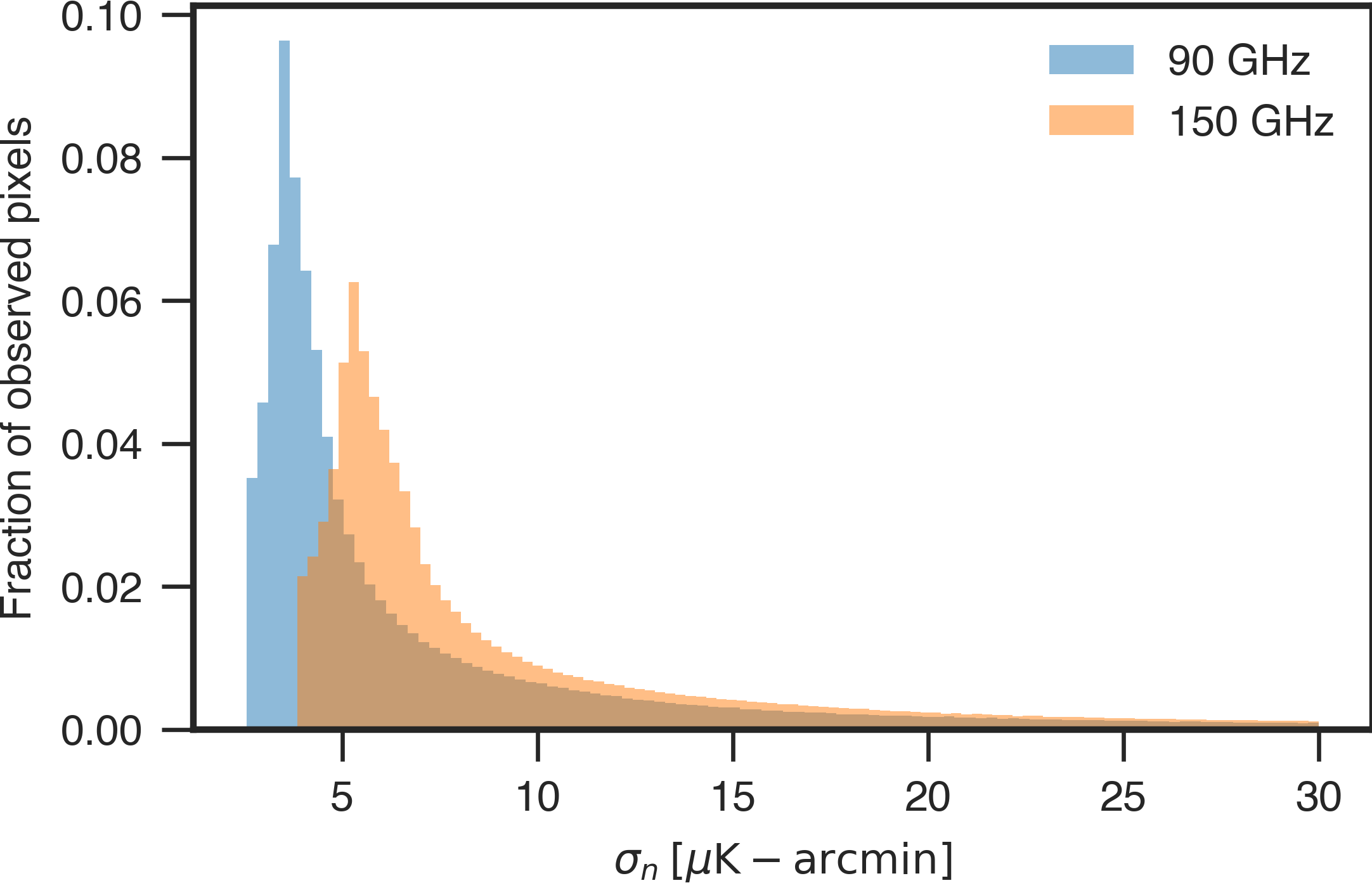}
    \caption{Simulated noise level distribution of AliCPT-1 90\,GHz (top left) and 150\,GHz (top right) in units of $\mu$K-arcmin. The two figures are in equatorial coordinate system, with a rotation of $[180^\circ,30^\circ]$, with a color scale adopting histogram normalization. The minimum and maximum are set to be 0 and 40 $\mu$K-arcmin. The bottom figure shows the histogram of noise standard deviation for the observed pixels with noise $\leq 30 \text{ }\mu$K-arcmin. The fraction of observed pixel is calculated with respect to the total number of observed pixels.}
    \label{fig:pol_noise}
\end{figure}

In the simulation, we adopt the sidereal fixed scanning strategy, centering at $\mathrm{RA}=180^\circ, \mathrm{DEC=30^\circ}$ in the northern sky, covering about $17\%$ of the sky area. The observing season is from October, 1st to next March, 31st.
In order to evaluate the noise level of AliCPT-1, we use MERRA2 reanalysis data \citep{MERRA2} from 2011 to 2017 to analyze the water vapor, pressure, temperature and other meteorological properties at the Ali site.
For each day of the observing season, we pick one day's data from the same dates over six years at random.
These mock meteorological data, combined with the instrumental attributes of AliCPT-1, are converted into the time streams of the detectors' noise variance. In Fig.\ref{fig:pol_noise}, we plot the map depth of AliCPT-1 scanning in the northern sky after one observation season. {Fig.~\ref{fig:pol_noise} also shows the histogram of the noise standard deviation of the observed pixels with noise level $\leq 30$ $\mu$K-arcmin.}

Two kinds of filters, polynomial and ground subtraction, are used in the time domain during simulation.
We perform a third-order polynomial fit on each detector's data and subtract it for each uniform azimuthal rotation.
To perform ground subtraction, for each detector, we project about half an hour's data into the horizontal coordinate system to get the ground template which is removed from the original data. 
These two time domain filters not only subtract the large scale signal of the atmosphere emission and the ground radiation, but also affect the extraterrestrial CMB and foreground emissions.
To study the loss of data and its recovery, the two filters can be combined into four cases by whether they are enabled.

In addition to the data of 90\,GHz, 150\,GHz bands of AliCPT-1, in order to increase the frequency coverage and improve the efficiency of foreground removal, we also generated `re-observed' map sets using sky maps from Planck HFI four bands and WMAP K band. The CMB and foreground sky maps are simulated by the Planck Sky Model (PSM) \citep{PSM:2013}, noise maps of Planck HFI are obtained from Planck Archive Legacy, and those of WMAP K band are generated with its noise covariance matrix. We re-observe the coadded sky map consisting of signal and noise in each band, using the same number of detectors as AliCPT-1's single band. We apply the same filters to the time-ordered data (TOD), and obtain the re-observed map at the end of the map-making procedure.

The data product obtained from the simulation pipeline can be divided into two categories: the Simulated data and the Ancillary data.
\begin{itemize}
\item Simulation data: 

    Simulation data only contain one set of observation/re-observation maps in all seven bands, including two AliCPT-1 bands, four Planck HFI and WMAP-K bands. Each map consists of CMB radiation, foreground emission and the random noise as follows.

\begin{itemize}
\item CMB maps.  
    We set the value of each cosmological parameter, which is totally blind to the analysts, as the Planck 2018's best-fit value adding a random fluctuation within its standard deviation. Specifically the tensor-to-scalar ratio equals to 0.023. Note that, throughout this paper, we assume the PGWs with the spectral index $n_t=0$ \footnote{Note that, throughout this work, we did not consider the lensed $B$-mode polarization in the simulation. In this article we focus on the constraint on PGWs and the lensed $B$ mode can be treated as a kind of noise with known spectrum, which is small in the multipole range used in this analysis. Therefore, we anticipate the lensed $B$ mode should not significantly influence the results presented in this paper.}.
\item Foreground maps.
     In the simulation, synchrotron, free-free, anomalous dust emission (AME), {here assumed to be due} to spinning dust grains, thermal dust emission, and the carbon monoxide emission signals are taken into account. The galactic model is set to simulation mode, so that the fluctuations at small scale are generated at random. Strong and faint point sources are also included in the simulation.
\item Noise maps.
     We obtain AliCPT-1's noise maps by projecting six month's noise TODs. For Planck HFI re-observations' noise maps we propagate the 87th FFP10 noise simulation through the re-observation pipeline, and for WMAP-K band we generate one specific noise realization based on the covariance matrix map.
     
\end{itemize}
\item Ancillary data:
\begin{itemize}
\item White noise covariance matrix of AliCPT-1 dual-band;
\item 50 observation seasons, re-observed CMB maps of all bands $\times$ four filter combinations.
       We set the values of the cosmological parameters as the best-fit values of Planck 2018 result, except, for these ancillary data sets, $r$ which is set to a value of 0.01 for all maps;
\item 50 observation seasons, re-observed CMB maps of all bands $\times$ four filter combinations, for $r=0$;
\item 100 observation seasons, noise maps of AliCPT-1 dual-band (based on meterological data) $\times$ four filter combinations;
\item 50 observation seasons, re-observed noise maps of Planck HFI 4 band (based on the FFP10 noise simulations of the Planck Legacy Archive) $\times$ four filter combinations;
\item 50 observation seasons, re-observed noise maps of WMAP K band (based on WMAP noise covariance matrix) $\times$ four filter combinations;
\item 2 realizations, re-observed foreground maps of all bands $\times$ four filter combinations.
\end{itemize}
\end{itemize}

The map sets listed above were used to do what we call the data challenge 1 (DC1). 
Participants try to estimate the power spectra of the Simulated Data and the corresponding cosmological parameters, with the help of the Ancillary Data. With the data challenge, we can test the data processing pipeline and get the forecast of the telescope performance. In addition, another map set was generated by subtracting the CMB maps from the Simulated Data in order to perform a null test and estimate the contribution and biases originating from the noise and foreground in our results.

\section{Estimators for $B$-mode power spectrum}
\label{sec:Cl_estimators}
CMB polarization can be completely characterized by spin ($\pm 2$) fields, denoted $P_{\pm}$. In terms of the $Q$ and $U$ Stokes parameters (defined in direction $\hat{n}$ with respect to a spherical coordinate system $\hat{e}_\theta, \hat{e}_\phi$), the two polarization fields are given by $P_{\pm}\equiv (Q \pm iU)$. The multipole decompositions of polarization are given by \cite{seljak1996},
\begin{equation}
P_{\pm}(\hat{n})=-\sum_{\ell m} (a^E_{\ell m} \pm a^B_{\ell m})~{}_{\pm 2}Y_{\ell m}(\hat{n}).
\end{equation}
In the partial sky case, the polarization field cannot be uniquely decomposed into $E$ and $B$ modes due to the ambiguity in the relationship between the Stokes parameters and various methods have been proposed to alleviate the partial sky $E$-to-$B$ mixing problem \cite{2002ApJ...567....2H, 2003MNRAS.343..559H, smith2006pseudo, 2007PhRvD..76d3001S, 2010PhRvD..82b3001Z, 2010A&A...519A.104K, 2011A&A...531A..32K, grain2012, 2019PhRvD.100b3538L,Ghosh2020}. In this article, we adopt the pure $B$-type polarization method, also called the Smith-Zaldarriaga (SZ) method, \cite{smith2006pseudo,2007PhRvD..76d3001S,chen1,chen2}), and a novel template cleaning (TC) method \cite{2019PhRvD.100b3538L}, to reconstruct the $B$-mode power spectrum for AliCPT. We summarize the pseudo-$C_\ell$ (PCL) estimators for these methods below.

\subsection{PCL-SZ estimator}
\label{ssec:SZ_estimator}
The key of SZ method is to construct two mutually orthogonal scalar (pseudo-scalar) fields $\mathcal{E}$ and $\mathcal{B}$. For an incomplete sky observation case, we can define the partial-sky harmonic coefficients of the $\mathcal E$- and $\mathcal B$-fields (with overhead tilde) as \citep{efstathiou2004},
\begin{eqnarray}
\tilde{\mathcal E}_{\ell m} &=& -\frac{1}{2}\int d\hat{n} \bigg\{P_{+}(\hat n)\left[\bar\eth \bar\eth\left(W(\hat{n})Y_{\ell m}(\hat{n})\right)\right]^\ast    +P_-(\hat n)\left[\eth\eth\left(W(\hat{n})Y_{\ell m}(\hat{n})\right)\right]^\ast \bigg\}, \label{pureee}\\
\tilde{\mathcal B}_{\ell m} &=& -\frac{1}{2i}\int d\hat{n}\bigg\{P_+(\hat n)\left[\bar\eth\bar\eth\left(W(\hat{n})Y_{\ell m}(\hat{n})\right)\right]^\ast   -P_-(\hat n)\left[\eth\eth\left(W(\hat{n})Y_{\ell m}(\hat{n})\right)\right]^\ast \bigg\} \label{purebb},
\end{eqnarray}
where $W(\hat{n})$ is the window function. These expressions can be expanded and simplified further for implementation and full form expressions can be found in \cite{2016RAA....16...59W}.

The partial-sky pure harmonic coefficients $\tilde{\mathcal E}_{\ell m}$ and $\tilde{\mathcal B}_{\ell m}$ are related to the full-sky harmonic coefficients $E_{\ell m}$ and $B_{\ell m}$ as follows,
\begin{eqnarray}
\tilde{\mathcal E}_{\ell m} &=&\sum_{\ell'm'}[K^{EE}_{\ell m,\ell'm'}E_{\ell'm'}+i K^{EB}_{\ell m,\ell'm'}B_{\ell'm'}], \label{Elm}\\
\tilde{\mathcal B}_{\ell m} &=&\sum_{\ell'm'}[-i K^{BE}_{\ell m,\ell'm'}E_{\ell 'm'}+K^{BB}_{\ell m,\ell'm'}E_{\ell'm'}], \label{Blm}
\end{eqnarray}
where $K^{ru}_{\ell' m' \ell m}$, with $r, u \in \{E,\ B\}$, represents the mixing kernels of pure fields. And then we can define pseudo estimators of $\mathcal{E}$- and $\mathcal{B}$- fields as,
\begin{eqnarray}
\tilde{\mathcal{C}}_{\ell}^{\mathcal{EE}} \equiv \frac{1}{2l+1}\sum_m  \tilde{\mathcal{E}}_{\ell m} \tilde{\mathcal{E}}^*_{\ell m}; \qquad
\tilde{\mathcal{C}}_{\ell}^{\mathcal{BB}}  \equiv \frac{1}{2l+1}\sum_m  \tilde{\mathcal{B}}_{\ell m} \tilde{\mathcal{B}}^*_{\ell m}.
\end{eqnarray}
They relate to the CMB power spectra $C_{\ell}^{EE}$ and $C_{\ell}^{BB}$ by the following relations
\begin{eqnarray}
\begin{pmatrix}   
\langle\tilde{C}^{\mathcal{EE}}_{\ell '}\rangle \\ \langle\tilde{C}^{\mathcal{BB}}_{\ell '} \rangle
 \end{pmatrix}=\begin{pmatrix} \mathcal{M}^{EE}_{\ell \ell '} & \mathcal{M}^{EB}_{\ell \ell '} \\  \mathcal{M}^{BE}_{\ell \ell '} &  \mathcal{M}^{BB}_{\ell \ell '} \\ \end{pmatrix} \begin{pmatrix}  
 {C}^{EE}_{\ell}  \\ C^{BB}_{\ell} 
 \end{pmatrix},
\label{SPCL-estimator30}
\end{eqnarray}
where the notation $\langle\cdot\cdot\cdot\rangle$ denotes the ensemble average. The exact expressions for the  mixing kernels $K^{ru}_{\ell' m' \ell m}$ and mixing matrices $\mathcal{M}^{ru}_{\ell \ell'}$ can be found in \cite{2009PhRvD..79l3515G,ferte2013}. 
Therefore, we can construct the unbiased estimators of the actual power spectrum as follows,
\begin{eqnarray}
\begin{pmatrix}   \hat{C}^{EE}_{\ell}  \\ \hat{C}^{BB}_{\ell}  \end{pmatrix}=\begin{pmatrix} \mathcal{M}^{EE}_{\ell \ell '} & \mathcal{M}^{EB}_{\ell \ell '} \\  \mathcal{M}^{BE}_{\ell \ell '} &  \mathcal{M}^{BB}_{\ell \ell '} \\ \end{pmatrix}^{-1} \begin{pmatrix}  \tilde{C}^{\mathcal{EE}}_{\ell '} \\ \tilde{C}^{\mathcal{BB}}_{\ell '}  \end{pmatrix}.
\label{SPCL-estimator3}
\end{eqnarray}
This estimator is adopted in the McMfL pipeline. A slight variation to the PCL-SZ estimator is to construct pure-$B$ modes but not pure-$E$ modes. Then we can get a mixing matrix relating the partial sky $\begin{pmatrix} \tilde C_\ell^{EE} & \tilde{C}^{\mathcal{BB}}_{\ell} \end{pmatrix}^t$ and the actual power spectrum $\begin{pmatrix} \hat C_\ell^{EE} & \hat C^{BB}_\ell \end{pmatrix}^t$. This variant of the PCL-SZ estimator has been used in the ABS and the TF pipelines. In this work we use the PCL-SZ method implemented in python package of the NaMaster \cite{2019MNRAS.484.4127A}.
\subsection{PCL-TC estimator}
\label{ssec:TC_estimator}
In this work, we also adopt a novel estimator to reconstruct $B$-mode power spectrum form $Q$, $U$ maps, called the PCL-TC estimator. This method uses the template cleaning (TC) method proposed in \cite{2019JCAP...04..046L, 2019PhRvD.100b3538L,chen1} to obtain a leakage free $B$-mode map. The main idea of the TC method is to use the high-precision $E$-mode signal to construct the $E$-to-$B$ leakage template, and then subtract a linear fit of this leakage template from the original $B$-mode signal. This gives us the $B$-mode map in our observation window without the partial sky $E$-to-$B$ leakage.
    
    
    

Since the cleaned $B$ map is essentially the isolated $B$-mode pseudo-scalar field, we can apply the scalar PCL estimator to reconstruct its power spectrum. For the cleaned $B$ map, the pseudo-multipoles are defined as
\begin{eqnarray}
    \tilde B_{\ell m} &=& \int  B(\hat n)\, W(\hat n)\, Y^*_{\ell m}d\hat{n}.
\end{eqnarray}
From pseudo-multipole coefficients, we can define the following PCL power spectrum estimate
\begin{eqnarray}
    \tilde{C}^{BB}_\ell =\frac{1}{2\ell + 1}\sum_{m} \tilde B_{\ell m}\tilde B_{\ell m}.
\end{eqnarray}
By inverting the mode-mode coupling matrix between pseudo power spectrum and CMB power spectrum, we can built an unbiased estimator of CMB $B$-mode power spectrum
\begin{eqnarray}
 \hat{C}^{BB}_\ell = \sum_{\ell'}M^{-1}_{\ell \ell'} \tilde{C}^{BB}_\ell,
\end{eqnarray}
where
\begin{eqnarray}
M_{\ell \ell'}=\frac{2\ell'+1}{4\pi}\sum_{\ell''}(2\ell''+1)\begin{pmatrix}
\ell' & \ell'' & \ell \\ 0 & 0 & 0
\end{pmatrix}^2w^2_{\ell''}.
\end{eqnarray}
Here $w_{\ell}$ is the power spectrum of the window function $W(\hat{n})$. The PCL-TC estimator is used in the GLS and cILC pipelines. The scalar PCL is implemented with NaMaster.


\section{$B$-mode filter correction}
\label{sec:filter}

As mentioned in Sec. \ref{sec:mock_data}, the AliCPT timestream are filtered to remove atmospheric emission and ground pickup. This filtering has two effects on the recovered $B$-modes. {First, there is a scale dependent loss of $B$-mode power.} Second, there is a leakage from $E$-to-$B$ mode. These filtering effects are discussed in detail in Paper II. Hence the power spectra estimates obtained with the PCL estimators in Sec. \ref{sec:Cl_estimators} are further corrected to remove the effects of TOD filtering. Here, we will summarize our methods to deal with this.
\begin{figure}
    \centering
    \includegraphics[width=1\textwidth]{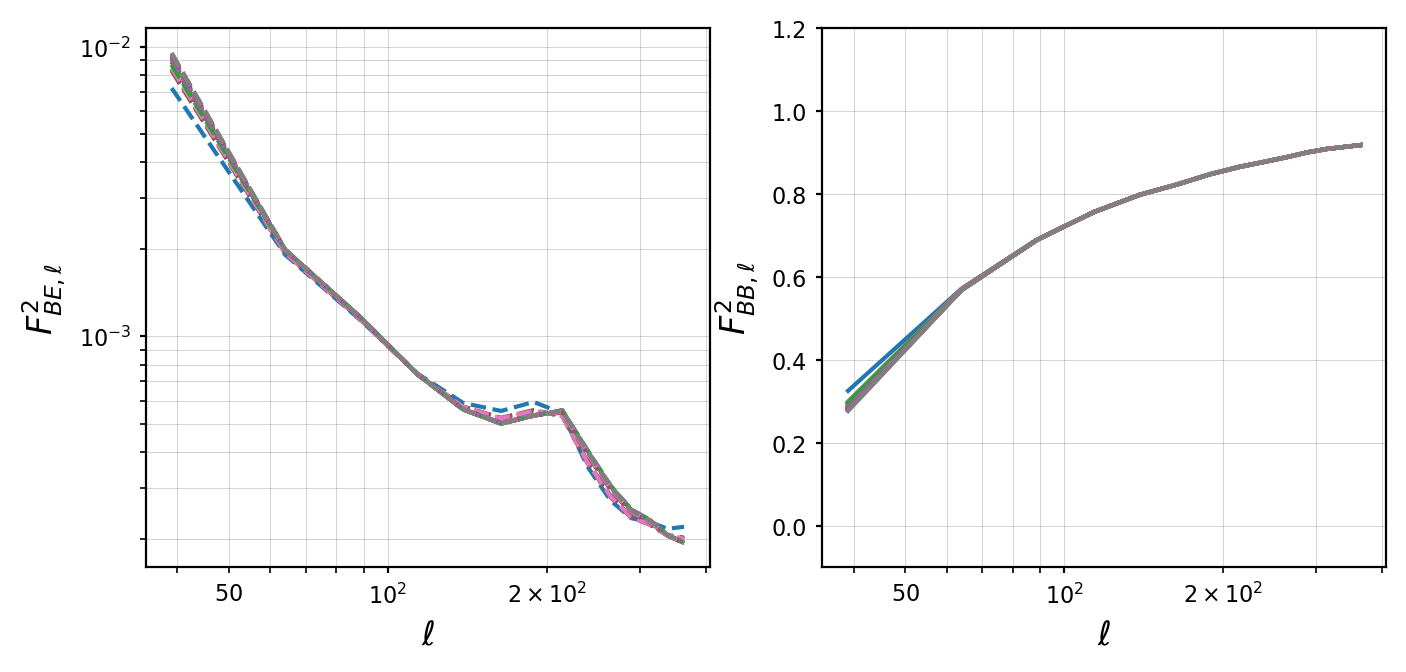}
        \caption{Plot of $F_{BE,\ell}^2$ (left) and $F_{BB,\ell}^2$ (right)  for the McMfL pipeline for all possible $\nu_1 \times \nu_2$ combinations. 
        Aside for small differences for $\ell < 40$, the transfer function mean is nearly identical for all frequency pairs.}
    \label{mcmfl-suppfac}
\end{figure}
If we consider a filtered CMB map, the $B$-mode power spectrum obtained from it, $D^{BB}_{\ell, \text{ filt}}$, will consist of two parts:
\begin{equation}
    D^{BB}_{\ell, \text{ filt}} = \bar D^{BB}_{\ell,S} + D^{BB}_{\ell,L},
\end{equation}
 where overhead bar indicates {power loss or power `suppression'} due to filtering, so $\bar D^{BB}_{\ell,S}$ is the suppressed $B$-mode power, and $D^{BB}_{\ell,L}$ is the power that has `leaked' to the $B$ mode from $E$ mode due to filtering. Note that throughout the paper we use $D_\ell$ to denote $\ell (\ell + 1) C_\ell/2\pi$. {Since the filtering process is linear, maps with multiple components, after filtering should be considered as the sum of the suppression and the leakage terms for each of the components.} 
 
\subsection{Transfer functions}
\label{ssec:transfer_fcn}
This filtered $B$-mode power can be written in terms of unfiltered $E$- and $B$-mode power as:
\begin{equation}
    D^{BB}_{\ell, \text{ filt}} = F_{BB,\ell}^2 D_\ell^{BB} + F_{BE,\ell}^2 D_{\ell, \text{ filt}}^{EE}.
\end{equation}
Here $F_{BB,\ell}^2$ and $F_{BE,\ell}^2$ represent the auto and cross-transfer functions. The auto transfer is responsible for power suppression and the cross transfer is responsible for leakage.
The transfer functions are computed from 50 filtered CMB only simulations. The transfer functions are computed as:
\begin{equation}
    \langle F^2_{BB, \ell} \rangle = \left\langle \frac{\bar D^{BB}_{\ell,S}}{D^{BB}_\ell}\right\rangle \qquad \langle F^2_{BE, \ell} \rangle = \left\langle \frac{D^{BB}_{\ell,L}}{D^{EE}_{\ell, \text{ filt}}} \right \rangle.
\end{equation}
\begin{figure}
    \centering
    \includegraphics[width=0.60\textwidth]{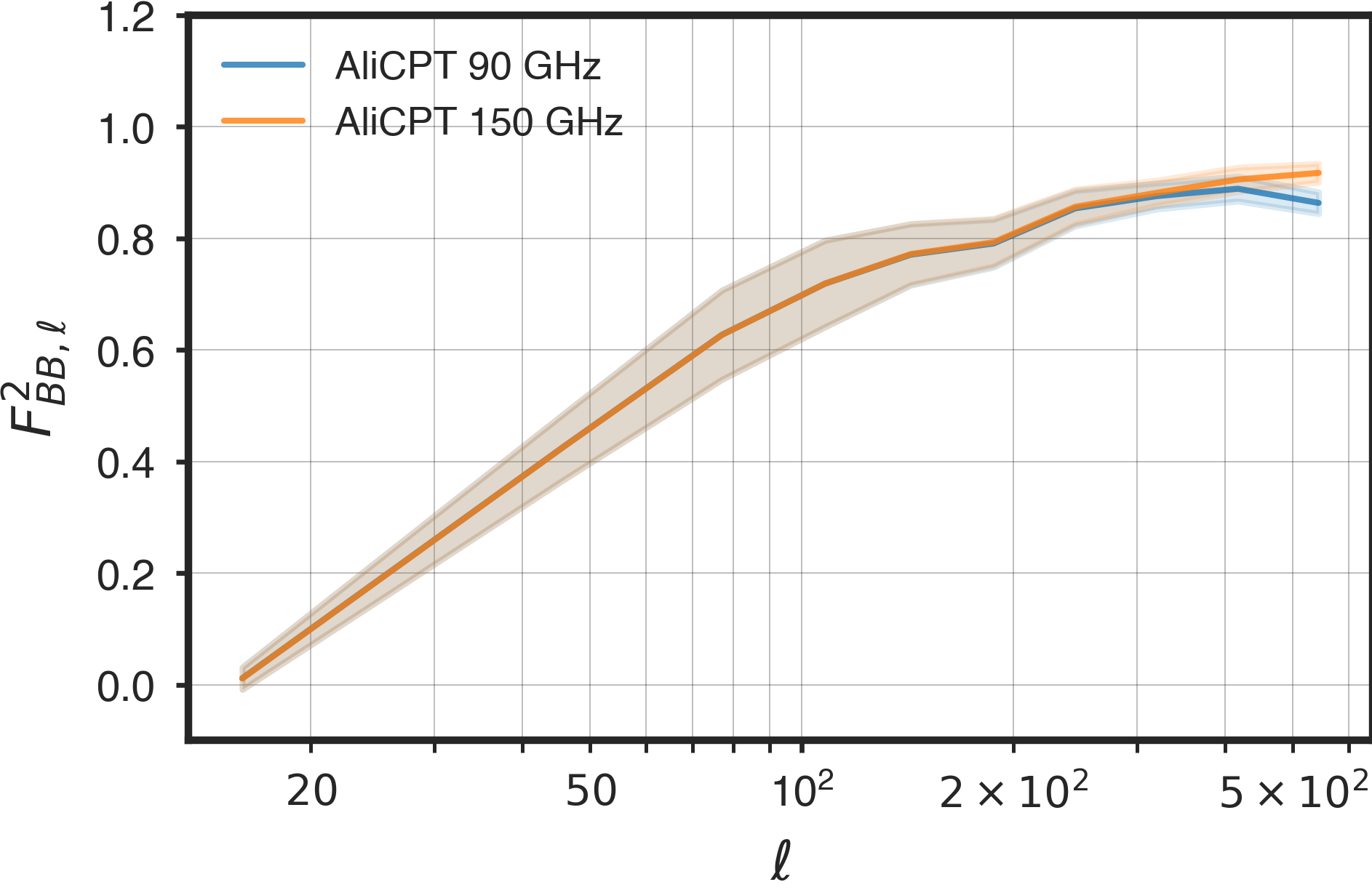}
    \includegraphics[width=0.37\textwidth]{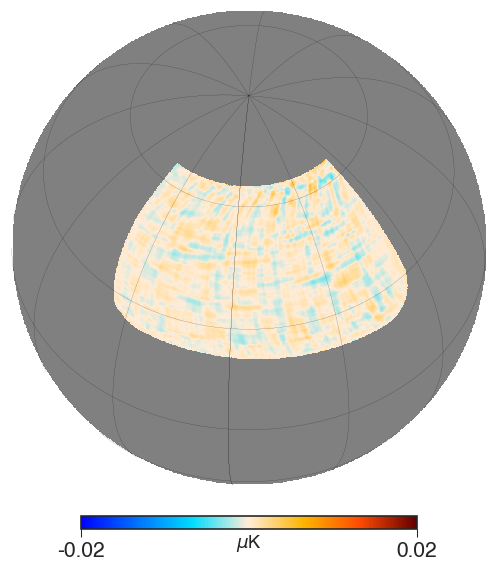}
    \caption{In the left plot we show the auto transfer function $F_{BB, \ell}^2$ for the two AliCPT channels used for GLS/cILC pipeline for correcting power suppression. Note that the transfer functions overlap in the multipole range of our interest. In the right plot we show the leakage template obtained from $T$- and $E$-mode information of WMAP K band and HFI 100, 143, 217, 353 GHz bands. The template shown here is apodized with the same inverse noise variance mask used in the GLS and cILC pipelines.}
    \label{fig:Planck-filter_corr}
\end{figure}
To compute the transfer functions, the CMB simulations from ancillary data discussed in Sec. \ref{sec:mock_data} are used. In this work, two different computation approaches are adopted. For the McMfL pipeline, all the power spectra calculated from the maps are using the PCL-SZ method. The leakage power 
($D^{BB}_{\ell,L}$) and filtered $E$-mode power spectrum ($D^{EE}_{\ell, \text{ filt}}$), are computed from $T$- and $E$-only filtered CMB maps. 
{The power suppressed $B$-mode spectrum $\bar D^{BB}_{\ell,S}$ is computed by subtracting  $\bar D^{BB}_{\ell,L}$ from the $B$-mode spectrum of the filtered CMB maps.} 
The unfiltered $B$-mode spectrum, $D^{BB}_\ell$ is computed from {the corresponding} unfiltered CMB {realization}. For all cases the BPSapo mask described in Sec. \ref{app:mask_apo} is used. In Fig. \ref{mcmfl-suppfac} we show the mean auto- and cross-transfer function used in the McMfL pipeline for all frequency channel pairs. We find some difference only for $\ell < 40$. 

The GLS and cILC pipelines do not use $F^2_{BE, \ell}$ to correct for the $E$-to-$B$ leakage. The leakage correction used in these pipelines is discussed in the next section. For computing $F^2_{BB, \ell}$ for GLS and cILC, we compute $\bar D^{BB}_{\ell,S}$ from $B$-mode only filtered $QU$ map using PCL-TC method with the inverse noise weighted mask, UNPinv discussed in Sec. \ref{app:mask_apo}. The unfiltered spectra $D^{BB}_\ell$ are computed from the unfiltered full-sky input map.
The full-sky power spectrum is computed with \texttt{anafast} function from \texttt{healpy} and binned with the same binning used in PCL-TC pipeline. The transfer functions are computed for each pair of input full sky and filtered CMB maps and use the mean transfer function from 50 simulations in our power spectrum correction. In the left plot of Fig. \ref{fig:Planck-filter_corr} we show the mean and standard deviation of the auto transfer functions $F^2_{BB, \ell}$ for AliCPT 90 GHz and 150 GHz channels. In the multipole range of our interest $F^2_{BB, \ell}$ is identical for all channels. {The small number of simulations used in these calculations can impact the estimates for the transfer functions. However, we expect the mean estimates of the transfer function to converge fairly soon. In future studies, we intend to repeat the calculations with significant increase in number of simulations.}

\subsection{Leakage correction with Planck}
\label{ssec:leakage_w_template}
Correcting the leakage using $F^2_{BE, \ell}$ is the standard process of correcting leakage. We consider a novel option for filtering leakage corrections using prior information from Planck $E$-maps to construct a template of the leakage. In the case of the GLS and the cILC pipelines where we obtain a foreground cleaned $B$-map, in harmonic space we can write:
\begin{equation}
    B^{\rm GLS/cILC}_{\ell m, \text{ filt}} = \bar B^s_{\ell m} + \bar B^n_{\ell m} + \bar B^f_{\ell m} + L^s_{\ell m} +  L^{n \prime}_{\ell m},
    \label{eq:cleaned_lm}
\end{equation}
where $B^{\rm GLS/cILC}_{\ell m, \text{ filt}}$ is the GLS/cILC cleaned filtered $B$-mode harmonic coefficient, $\bar B^s_{\ell m}$ is the suppressed $B$ modes {of the CMB} signal, $\bar B^n_{\ell m}$ is the suppressed residual noise in the cleaned $B$-mode, 
$\bar B^f_{\ell m}$ is the suppressed residual foreground in the cleaned $B$-mode, $L^s_{\ell m}$ is the leakage term of the CMB $E$ modes due to filtering and $L^{n\prime}_{\ell m}$ is the projected leakage of the $E$-mode noise. For our analysis the main contribution to the cleaned $B$-map comes from the AliCPT channel, implying $L^{n \prime}_{\ell m} < L^s_{\ell m}$, as AliCPT 90 and 150 GHz noise level is smaller than the $E$-mode CMB signal. Also note that $B^{\rm GLS/cILC}_{\ell m, \text{ filt}}$ should pick up a leakage contribution from the $E$-mode foregrounds, but this should be removed by foreground cleaning methods as it will have different frequency scaling compared with the CMB. So any residual leakage from foreground should be small.

The Planck experiment measures $E$-modes which are independent of filtering effects of AliCPT. The Planck $E$-mode maps can therefore be propagated through the AliCPT observation pipeline to be `reconstructed' to get the $B$-mode leakage template $\boldsymbol{L}$. First we obtain the NILC cleaned $T$- and $E$-modes from Planck HFI and WMAP K band combination. Then we get the $IQU$ map from these $T$- and $E$-modes. This $IQU$ map is then `reconstructed' by propagating through the AliCPT simulation pipeline. The $B$-map obtained from the filtered $T$ and $E$ mode only $IQU$ map is the leakage template. The leakage template used in the DC1 analysis is shown in the right plot of Fig. \ref{fig:Planck-filter_corr}. Note that Planck $E$-modes have much larger noise contamination than AliCPT. The filtering effect is linear, so in harmonic space:
\begin{equation} 
    L_{\ell m} = L^s_{\ell m} + L^n_{\ell m},
    \label{eq:leak_lm}
\end{equation}
where the spherical harmonic coefficients $L_{\ell m}$ of the template is a sum of the contribution from the $E$-mode signal $L^s_{\ell m}$, and the $E$-mode noise $L^n_{\ell m}$. Since we have filtered the NILC cleaned maps to construct the template, the leakage of any residual $E$-mode foreground should be very small. The noise leakage term $L^n_{\ell m}$ is different from $L^{n\prime}_{\ell m}$ in Eq. \eqref{eq:cleaned_lm}, as the dominant contribution in the later is from the AliCPT channels which are not used in constructing the leakage template. Finally the cross spectrum of $\boldsymbol B^{\rm GLS/cILC}_\text{filt}$ and $\boldsymbol L$ will give us the leakage bias at power spectrum level:
\begin{equation}
    C^{L \text{ bias}}_\ell = \langle L_{\ell m} (B^{\rm GLS/cILC}_{\ell' m', \text{ filt}})^*\rangle \simeq \langle L^s_{\ell m} (L^s_{\ell' m'})^*\rangle.
    \label{eq:leak_bias}
\end{equation} 
There will be a small correlation between $L^n_{\ell m}$ and $L^{n\prime}_{\ell m}$ but we found it to be small as $L^{n\prime}_{\ell m}$ mostly gets contribution from the AliCPT channels which are absent in $L^n_{\ell m}$. We subtract $C_\ell^{L \text{ bias}}$ to correct for the $E$-to-$B$ filtering leakage. The contribution of $L^{n\prime}_{\ell m}$ to the final power spectrum is smaller than that of $L^s_{\ell m}$ and is corrected in the mean from the noise debiasing performed with filtered noise simulations. 
We debias the power spectrum obtained from the foreground cleaned maps for the noise bias and the leakage bias. Finally the power suppression is corrected by dividing the binned debiased power spectrum by the auto transfer functions, $F^2_{BB, \ell}$.
The Planck based correction method has been adopted in the GLS and cILC pipelines.

\subsection{Obtaining filtering corrected spectrum}
\label{ssec:filter_correct}
In general the power spectrum of a filtered observation map can be written as:
\begin{equation}
    D^{BB}_{\ell, \text{ filt}} = \sum_c \left[ \bar D^{BB}_{\ell,S, c} + D^{BB}_{\ell,L, c} \right] + N_{\ell, \text{ filt}}^{BB}, 
\end{equation}
where $c$ is the index over various emission components and $N_{\ell, \text{ filt}}^{BB}$ is the filtered $B$-mode noise power spectrum. In case of the GLS and cILC method we assume that the map contains only the CMB signal, but for McMfL it is CMB and foreground. The noise bias $N_{\ell, \text{ filt}}^{BB}$ is computed from the 50 noise simulations mentioned in Sec. \ref{sec:mock_data}. The estimate of the corrected $B$-mode spectrum is then given by:
\begin{equation}
    \hat D^{BB}_\ell = \frac{1}{F_{BB,\ell}^2}\left[ D^{BB}_{\ell, \text{ filt}} - N_{\ell, \text{ filt}}^{BB} - \sum_c D^{BB}_{\ell,L, c} \right].
    \label{eq:filter_corr}
\end{equation}
The leakage contribution $\sum_c D^{BB}_{\ell,L, c}$, for McMfL is computed as $F_{BE,\ell}^2 \sum_c D^{EE}_{\ell, c}$, using the $E$ modes of the CMB and foregrounds. {While, in the GLS or cILC pipeline we use $D_\ell^{L \text{ bias}}$ computed with the leakage template as $\sum_c D^{BB}_{\ell,L, c}$.} For the McMfL pipeline this procedure is applied to each $\nu_1 \times \nu_2$ 
spectra, but we do not show the frequency indices explicitly in Eq.~\eqref{eq:filter_corr}.

\subsection{Forward modeling of filtering effects}
\label{ssec:forward_model}
The filtering corrections using transfer function and cross spectra from leakage template represents a `backward correction process' to recover the corrected $B$-mode power spectrum. Alternatively, we can completely avoid the correction of the $B$-mode power spectrum in favor of the `forward modeling' of the filtering effects, in fitting the CMB $B$-mode parameters like $r$ and $A_{\rm lens}$. We use this `forward modeling' of filtering effects for the ABS and template-fitting parameter estimations. The method is discussed in detail in Paper III, and also is elaborated later in Sec.~\ref{ssec:abs}.

Now we will provide a brief summary of the forward modelling approach. The filtering effects are determined by a linear mapping of the filtering matrix. Thus, one can use a set of simulations to construct several templates for each independent component of the filtered CMB, including the tensor-only component, the $E$-to-$B$ leakage component, and the lensing one. The tensor-only component is built by the mean of the band powers by averaging over 50 realizations of `reconstructed' maps, each of them being generated with a fiducial value of $r = 0.01$ and entering into the observation pipeline. Thus, an arbitrary tensor-only band power can be achieved by simple rescaling such template by a factor of  $r/0.01$, so that by fitting the parameter $r$, one can directly match the template to data without correcting for filtering effects. In the same way, we can build for the $E$-to-$B$ leakage template by using 50 realization $TE$-only maps, which are generated by the fiducial $\Lambda$CDM model. 
The $E$-to-$B$ filtering leakage contribution in the $BB$ powers does not depend on $r$, and we can compute the leakage contribution using $E$-mode only simulations. We correct the leakage contribution by directly subtracting the $E$-to-$B$ leakage band powers from the data without any fitting procedure.


In the likelihood analysis of the ABS and the TF pipelines, based on the constructed templates, the simulation tests have validated this template-fitting approach, showing the same performance in the $r$ estimate as that of the `backward correction' approach. It should also be noted that during the TF procedures, the fitted foreground parameters in fact characterize the filtered foreground components.

\section{Analysis pipelines}
\label{sec:pipelines}


We have five analysis pipelines for estimation of the tensor-to-scalar ratio. The first three methods - ABS, GLS and cILC, perform foreground cleaning of the multi-frequency data to obtain foreground cleaned power spectrum. {The ABS method denoises the input maps by removing the noisy modes from the full covariance matrix, and then obtains an analytical solution for the CMB component. Both the GLS and cILC methods assume a three component mixing matrix, with the CMB, dust, and synchrotron. The GLS simply solves the linear system assuming three templates for the three components. In the cILC method we minimize the total variance subject to constraints that exclude the average dust and synchrotron components. All three methods obtains multipole weights to combine the different frequency bands and obtain the cleaned map.} The foreground cleaning aspect of these three methods have been discussed in detail in Paper III. We will briefly review the foreground cleaning of these three methods and then discuss in detail the tensor-to-scalar ratio estimation from the cleaned spectra. The other two methods, McMfL and TF, perform global fit of CMB \emph{and} foreground parameters from the multi-frequency data, {by modelling the power of the CMB and the foreground components. These two methods differ in their modelling of the foreground components, and their approach to filtering corrections. The choice of mask and power spectrum estimation methods are different for the different pipelines, with different choices of multipole ranges and bin widths. Due to these differences there are variations in the multipole range of analysis for the different pipelines.}

\subsection{ABS pipeline}
\label{ssec:abs}


The analytic blind separation (ABS) method is a novel, computationally efficient method for the blind separation of the CMB from foregrounds~\cite{2016arXiv160803707Z}, which has been tested extensively for extracting the CMB temperature and polarization $E$- and $B$-mode power spectra from the simulated maps~\cite{2018ApJS..239...36Y,2019arXiv190807862S}. The ABS estimator acts directly on measured multi-frequency cross band powers, which extracts CMB power spectra by projecting the data onto CMB subspace. In the presence of instrumental noise and a limited number of frequency channels, the ILC and ABS would yield slightly different results~\cite{Vio:2008us,2008PhRvD..78b3003S,2019arXiv190807862S}. 
{Assuming that the measured cross band power spectrum between the $i$- and $j$-th frequency channels in the multipole bin $\ell$ is given by $ D^{\rm obs}_{ij}(\ell)$,
the unique solution of the CMB in the noise-free case is achieved by the Sylvester’s determinant theorem and reads  
\beq\label{eq:abs}
\mathcal{D}^{\rm cmb}(\ell) = \left( \sum_{\mu=1}^{M+1} G^2_{\mu}\lambda_{\mu}^{-1}\right)^{-1}\,, {\rm with}\,\,G_\mu = {\bf f\cdot E}^\mu\,,
\eeq
as long as $M< N_f$, where $M\equiv rank\left(\mathcal{D}_{ij}^{\rm fore}(\ell)\right)$, and we choose $f_i=1$ for all channels so as to satisfy the emission law of the CMB in the units of thermodynamic temperature. ${\bf E}^\mu$ and $\lambda_\mu$ stand for the $\mu$-th eigenvector and associated eigenvalue of $\mathcal{D}^{\rm obs}_{ij}(\ell)$. We adopt the normalization condition for eigenvectors, ${\bf E}^\mu \cdot {\bf E}^{\nu} =\delta_{\mu \nu}$. This is similar to the idea of the harmonic-space ILC, in that the data are projected onto certain basis vectors to uniquely extract the CMB signal. }



{In the presence of instrumental noise, which is assumed to be an uncorrelated Gaussian distribution with zero mean and rms levels of $\sigma_{\mathcal{D},i}^{\rm noise}$ for the $i$-th frequency channel, $i=1,2\cdots N_f$, the noise-free solution of Eq.~\ref{eq:abs} then can be recast into 
\beq\label{eq:abs1}
\hat{D}^{\rm cmb} = \left( \sum^{\tilde{\lambda}_{\mu}\geq \tilde{\lambda}_{\rm cut}} \tilde{G}^2_{\mu}\tilde{\lambda}_{\mu}^{-1}\right)^{-1} - \mathcal{S},
\eeq
where 
\beq\label{eq:noiseD}
\tilde{D}^{\rm obs}_{ij}\equiv \frac{D^{\rm obs}_{ij}}{\sqrt{\sigma_{D,i}^{\rm noise}\sigma_{D,j}^{\rm noise}}} + \tilde{f}_i\tilde{f}_j\mathcal{S}, \text{   with  }
\tilde{f_i} \equiv \frac{f_i}{\sqrt{\sigma_{D,i}^{\rm noise}}}; \text{  and  }\tilde{G}_{\mu}\equiv {\bf \tilde{f}}\cdot {\bf \tilde{E}}^\mu.
\eeq
Here ${\bf \tilde{E}}^\mu$ and $\tilde{\lambda}_\mu$ are the $\mu$-th eigenvector and corresponding eigenvalue of $\tilde{D}^{\rm obs}_{ij}$, respectively. Note that in the above expressions we have chosen to not write the $\ell$ index explicitly. As seen,  the ABS method thresholds the eigenvalues to keep only signal dominated modes, and we choose $\tilde{\lambda}_{\rm cut} = 1$ here, as suggested in ~\cite{2018ApJS..239...36Y}. The one important essence of the ABS is to introduce a positive \emph{shift} parameter, $\mathcal{S}$, in Eq.~(\ref{eq:abs1}), corresponding to that we artificially add a positive CMB signal to the data and then subtract it out of the solution. This is particularly important for low signal-to-noise regime and responsible for stabilizing the computation as long as it is large enough. From tests, we choose $\mathcal{S}= 10\mu {\rm K}^2$ in our analysis.}

\begin{figure}[!h]
    \centering
    \includegraphics[width=0.65\textwidth,height=0.4\textwidth]{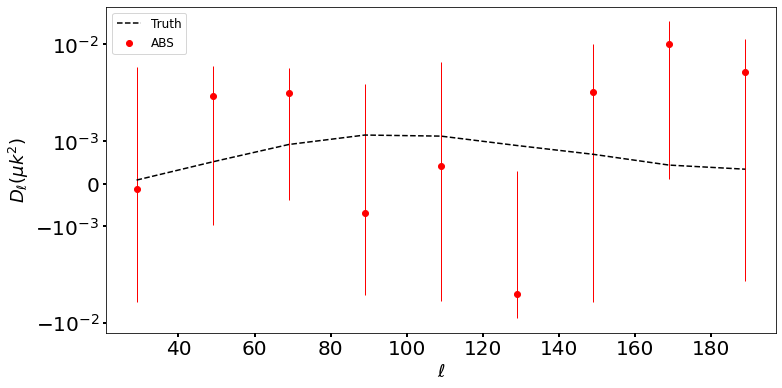}
    \caption{Comparison of the CMB $BB$ tensor-only band powers and associated $1\sigma$ error bars recovered by the ABS pipeline (red) and the simulation truth of $r=0.023$ (black dotted), in the range of $\ell\in [20,200]$, with 9 $\ell$-bins of $\Delta \ell=20$. Note that, due to the strong noise fluctuations, the recovered amplitudes appear negative in some $\ell$-bins. The $E$-to-$B$ leakage has been debiased in the data based on the leakage template, and the plots for both tensor-only signals shown here do not make any filtering corrections (such as the suppression effect), as the use of "forward modeling" in the ABS pipeline (see Sec.~\ref{sec:filter}).}
    \label{fig:ABS-fit}
\end{figure}


The seven frequency channel DC1 simulations are analyzed on a sky patch with $f_{\rm sky}\sim 5\%$, by adopting the ABSapo mask. Details of the mask is further discussed in Sec.~\ref{app:mask_apo} and shown in Fig.~\ref{fig:apo_mask}.
The plot in Fig. \ref{fig:ABS-fit} presents the ABS derived CMB $BB$ band powers in the range of $\ell \in [20,200]$ with 9 $\ell$-bins, where the filtering effects on both band powers are not corrected since we use forwarding modelling of filtering effects in the ABS pipeline as mentioned in Sec.~\ref{sec:filter}. Note that, the chi-squared value when fitting a model to data does not change if the model-predicted band powers and associated covariance have already incorporated filtering effects (refer to Paper III for details).

The estimation on tensor-to-scalar ratio $r$ through the ABS pipeline can be decomposed into two sequential steps. In short, we first recover the CMB band powers in each multipole bins through ABS estimator, and then we fit the recovered powers to the fiducial cosmological model to obtain the final estimate on $r$. The overall $BB$ band powers are modeled as a superposition of the tensor-only and the  $TE$-only components, which reads  
\beq\label{eq:cmbt} 
D^{BB}(\ell) =\frac{r}{0.01}\times \bar{D}^{BB}_{r=0.01}(\ell)+\bar{D}^{BB}_{\rm TE~only} \,.
\eeq
Here the tensor-only template $\bar{D}^{BB}_{r=0.01}$ consists merely of the filtered primordial $B$-mode signal, which is built from the mean of the band powers computed from 50 realization filtered CMB polarization maps with a fixed $r=0.01$, where the ensemble average of $BB$ spectrum from the $E$-to-$B$ leakage has been subtracted by using the leakage template, $\bar{D}^{BB}_{\rm TE~only}$. Such leakage template is constructed by the mean of $BB$ power spectra of 50 $TE$-only filtered CMB simulation maps, where the input CMB spectra are generated using the fiduical cosmological parameters but we explicitly null the $BB$ spectrum with setting $r=0$. All these maps used for building the templates are based on the ancillary data sets mentioned in Sec.~\ref{sec:mock_data}.



In the ABS analysis, we employ a Gaussian likelihood (defined in Eq.~\eqref{eq:gauss_like} and discussed in Sec.~\ref{app:likelihood})  to estimate the tensor-to-scale ratio, with using 9 $\ell$-bins in the multipole range $20\le \ell \le 200$. We stimate the covariance by using the ABS-derived band powers from 50 realization maps with the same generation process as for DC1 data, keeping the same foreground sky and the noise level as in DC1 data but setting the fiducial value of $r=0.01$ for $\Lambda$CDM model. These realizations are produced by combining the different component filtered maps from the ancillary data listed in Sec. \ref{sec:mock_data}. Since the CMB tensor contribution to the covariance is extremely minor compared with the noise and other contributions, the choice of fiducial value of $r$ has almost no changes on the estimated covariance, as long as it does not deviate greatly from the true value.  As a result, various uncertainties contributed from the noise, CMB signal, $E$-to-$B$ leakage and TOD filtering effects as well as foreground residuals after the ABS cleaning have been appropriately included in the covariance. 


An uniform prior on $r$ is assumed in the ABS analysis for $r\in [0,1]$. We sample the posterior distribution of the parameters using the Monte-Carlo Markov chain (MCMC) ensemble sampler \texttt{dynesty} \cite{1704.03459,1904.02180} with dynamic nested sampling scheme. As seen from Fig. \ref{fig:ABS-fit}, the recovered amplitudes of CMB $BB$ band power over all $\ell$-bins are almost consistent with the theoretical expectations within $1\sigma$ level, which can provide an accurate measurement of $r$ statistically. Note that, the band powers in the range of $80\le\ell\le140$ become negative, which is caused by large noise fluctuations in these bins as the mean noise subtraction has been applied to the data. We also find an increased statistical uncertainty as $\ell$ increases, which is mainly due to the noise band power goes like $\ell^2$. By combining all 9 $\ell$-bin measurements from DC1 data, we find the likelihood peaks at near the input and the best-fit value is $r=0.019$, which is listed in Table~\ref{tab:DC1_r_estimate_comparison}. The ABS likelihood yields the constraint, $r<0.082$ at 95\% confidence with $\sigma(r)=0.024$, fully consistent with the input $r=0.023$. The results for the null test will be shown in Table~\ref{tab:nulltest_r_95pcCL_comparison}, where in the absence of the primordial $B$-modes, bounds on $r$ can be derived, $r< 0.048$ at 68\% CL and $r < 0.069$ at 95\% CL (with a best-fit in zero), respectively. The normalized posterior for $r$ is shown in Fig.~\ref{fig:combined_posterior_plot} for both DC1 and null test cases.

\subsection{GLS pipeline}
\label{ssec:gls}

The generalised least squares (GLS) pipeline is a foreground cleaning pipeline where we assume our multi-frequency maps to be a linear mixture of three emission components, CMB, synchrotron and thermal dust. The cleaned $B$-mode map is obtained by solving the linear system. This method has been discussed in detail in Paper III. We only summarise the pipeline here.
For the GLS method, the observations, $\boldsymbol{d}(p)$, at pixel $p$, are modelled as:
\begin{align}
    \boldsymbol{d}(p) =& \boldsymbol{A s}(p) + \boldsymbol{n}(p). 
    \label{eq:model}
\end{align}
Here $\boldsymbol{A}$ denotes the mixing matrix of the three emission components, $\boldsymbol s$  represents the templates for the three components. The mixing matrix for the foreground components is computed by assuming a power-law emission for the synchrotron, and the thermal dust is modeled as a modified black body. We adopt Planck priors to compute the mixing matrix. The GLS pipeline assumes this linear system as a model for the observed sky and solves the linear system of Eq. (\ref{eq:model}) with inverse noise variance ($N$) weights. The GLS weights are given by \citep{Delabrouille2009}:
\begin{equation}
    \boldsymbol{W}^{\rm GLS} = \left[ \boldsymbol{A}^t \boldsymbol{N}^{-1} \boldsymbol{A} \right]^{-1}\boldsymbol{A}^t \boldsymbol{N}^{-1},
\end{equation}
and the template for component $c$ given by:
\begin{equation}
    \hat s_{c} = \sum_\nu W^{\rm GLS}_{c, \nu} d_\nu.
\end{equation}

\begin{figure}
    \centering
    \includegraphics[width=0.49\textwidth]{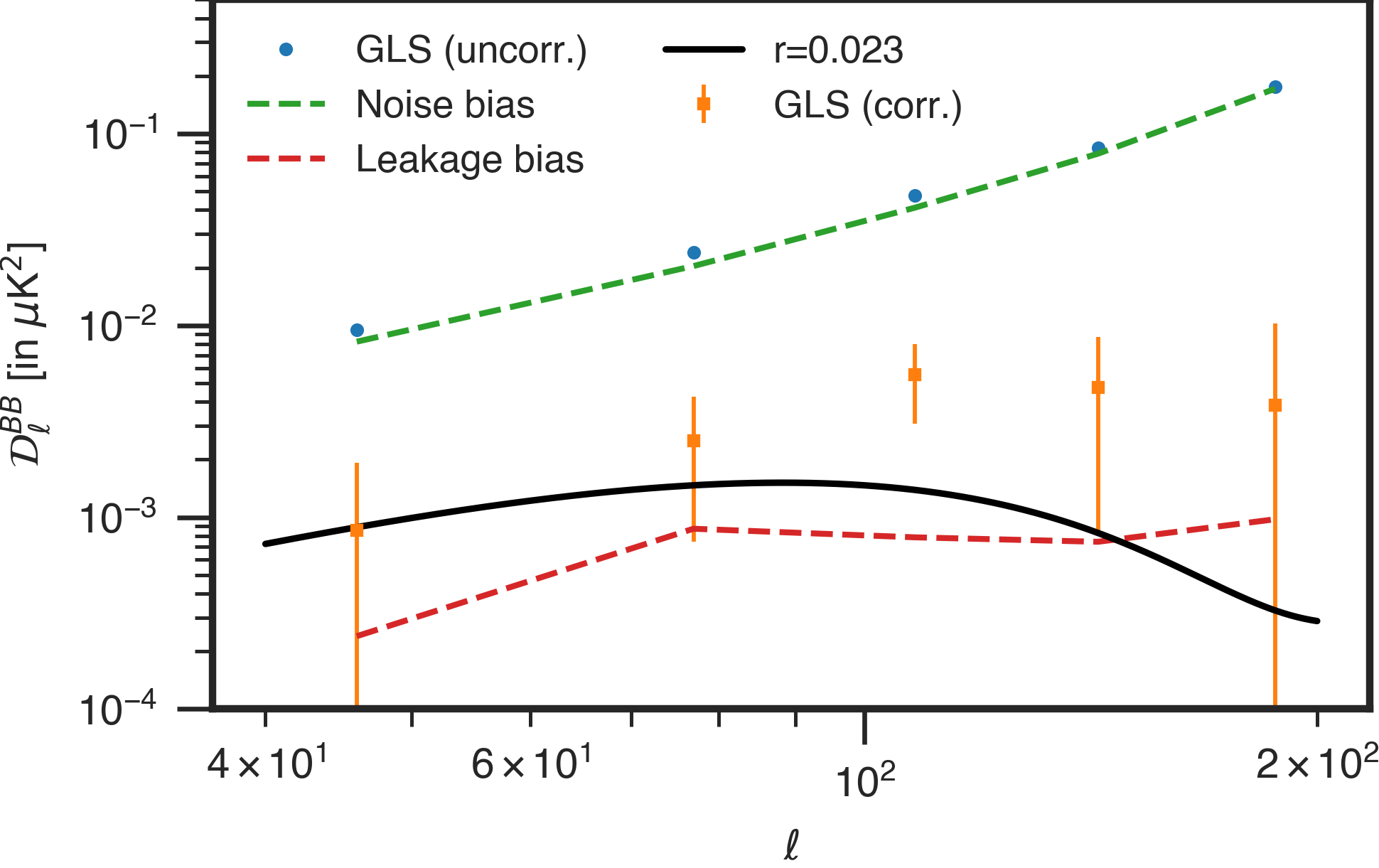}
    \includegraphics[width=0.49\textwidth]{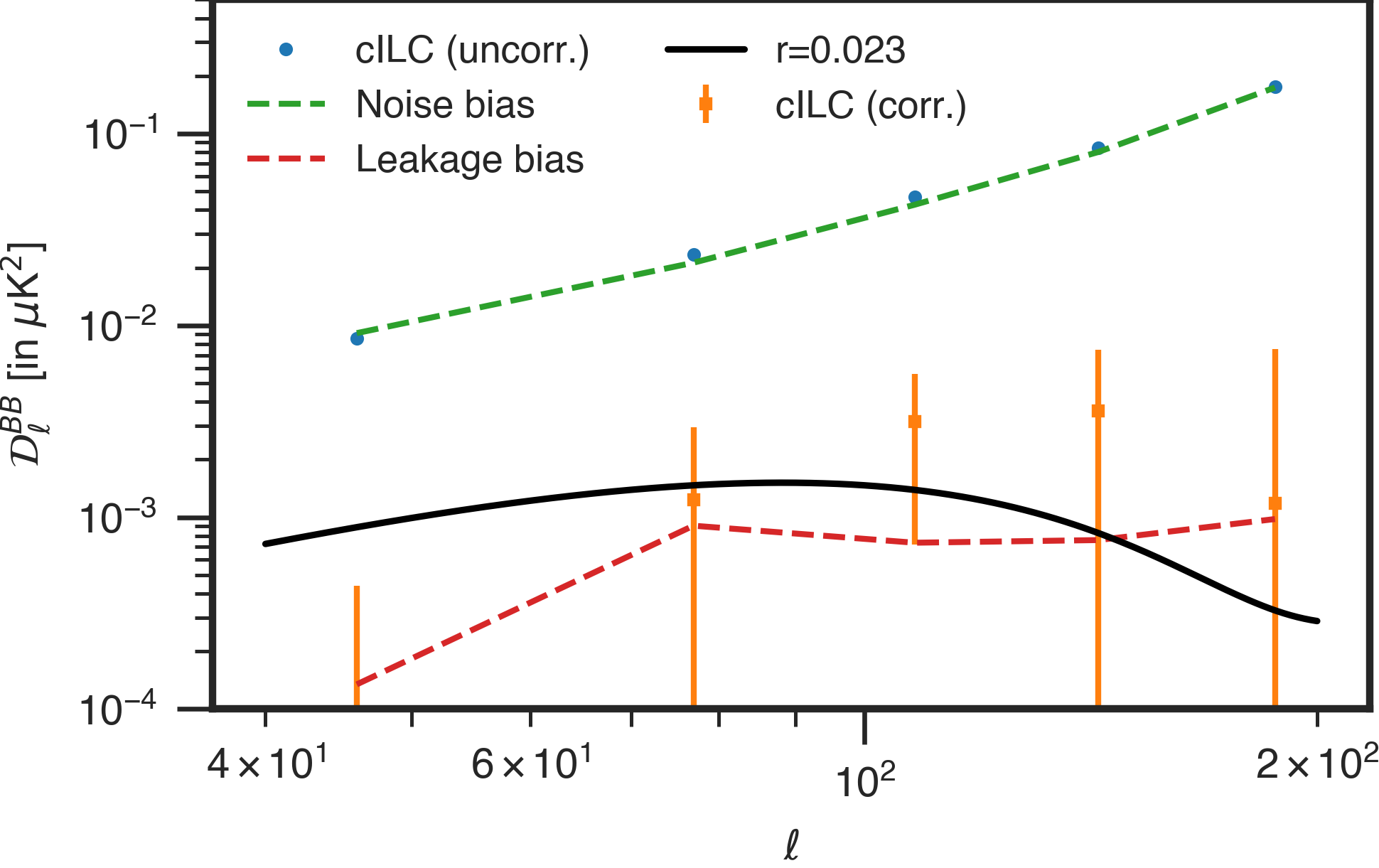}
    \caption{The power spectrum plots obtained after component separation with the GLS method (left) and the constrained ILC method (right). All spectra plotted here have been corrected by the auto-transfer function. The power spectrum estimated with PCL-TC without noise or leakage bias correction is shown by the blue markers. The noise bias is shown by the green dashed line, the leakage bias computed by taking cross spectrum with the leakage template is shown by the red dashed line. After debiasing for both noise and leakage bias we show the final $B$-mode spectra by orange markers. {The error bars are obtained from the diagonal part of the fiducial covariance matrix computed by propagating 300 unfiltered CMB $+$ noise $+$ foreground simulations through the GLS/cILC pipelines. This correctly includes the errors due to noise and residual foreground. However, it does not include errors arising from correction of the filtering process. We have estimated from filtered CMB $+$ noise simulations that we underestimate the error bars by few percent in the first bin, and its impact diminishes with multipole. In future work we will replace unfiltered simulations with filtered sets to correctly propagate all errors.}}
    \label{fig:GLS-cILC_PS}
\end{figure}

All DC1 bands are used in the analysis. The maps are pre-processed by masking for compact point sources and then interpolating in the masked spots. The UNPapo mask shown in Fig. \ref{fig:apo_mask} is used for further analysis of these maps. Refer to Sec.~\ref{app:mask_apo} and Paper III for point source pre-processing and mask choices. The GLS weights are computed and applied in harmonic space. The power spectrum is computed from the GLS cleaned CMB map with the PCL-TC estimator and debiased for noise and TOD filtering leakage. Finally we correct for power suppression due to the filtering by correcting with the transfer function. We show the power spectrum results for the GLS in Fig.~\ref{fig:GLS-cILC_PS}. For the parameter estimation we will use the power spectra estimates in multipole range 40-200.
 
We will use a Gaussian likelihood to fit for the value of $r$ and other model parameters $\theta$. The Gaussian likelihood has the form given in Eq.~\eqref{eq:gauss_like}. For the GLS pipeline, the theoretical power spectrum $C_\ell^{\rm th}$ is modeled as:
\begin{equation}\label{eq5.8}
    C_\ell^{\rm th}(r, A_d^{\rm res}) = \frac{r}{0.01} C^{\text{tens, }r=0.01}_\ell + A_{\rm lens} C^{\rm lensing}_\ell + A_d^{\rm res} C_\ell^{\rm dust},
\end{equation}
where we have introduced $A_d^{\rm res}$ as a nuisance parameter for the residual foreground contamination. For DC1, we use $A_{\rm lens}=0$ as these simulations are lensing free. The primary foreground contaminant for our experiment is thermal dust, so our foreground residuals only model the polarized dust emission. We adopt the model for residual dust power spectrum based on \cite{Planck2020:PolDust}:
\begin{equation}
    D_\ell^{\rm dust} = \sum_{\nu_1 \nu_2}W^{\rm GLS}_{{\rm CMB}, \nu_1, \ell}W^{\rm GLS}_{{\rm CMB}, \nu_2, \ell}\left(\frac{\ell}{80}\right)^{\alpha+2} \left(\frac{\nu_1 \nu_2}{353^2}\right)^{\beta_d-2}\frac{B_{\nu_1}(T_d)B_{\nu_2}(T_d)}{B_{353}(T_d)B_{353}(T_d)}\frac{C_{\nu_1}C_{\nu_2}}{U_{\nu_1}U_{\nu_2}}.
    \label{eq:dust_spectrum}
\end{equation}
Here $W^{\rm GLS}_{{\rm CMB}, \nu, \ell}$ denotes the GLS weights for the CMB in frequency channel $\nu$. The dust frequency scaling in Eq. \eqref{eq:dust_spectrum} is in brightness temperature units, while our observations are in thermodynamic temperature units. To convert the dust model to thermodynamic units we use a $C_\nu / U_\nu$ factor, where $C_{\nu}$ is the color correction factor to convert from dust spectrum to IRAS reference spectrum, $I_\nu = \nu^{-1}$, and $U_\nu$ is the unit conversion factor to convert thermodynamic temperature to brightness temperature with IRAS reference spectrum. Note that the $C_\nu / U_\nu$ factor is independent of the choice of reference spectrum. The color correction and unit conversion factors are calculated with bandpass integration following the prescription in \cite{Planck2014:HFIspec}. The GLS weights are multiplied to compute the residual dust power in the GLS CMB map. We set $\alpha = -2.54$, $\beta_d = 1.59$ and $T_d = 19.6$ K, from \cite{Planck2016:highLatDust, Planck2020:PolDust}. 

To compute the covariance matrix, $\boldsymbol{M}_{\rm fid}$, \tblue{defined in Sec.~\ref{app:likelihood}}, we use 300 simulations of CMB and noise, with fixed foreground simulations that include thermal dust, synchrotron and point sources. We set $r=0.03$ in the fiducial $C_{\ell}$s. The noise simulations use WMAP 9 year noise variance maps and for HFI we use FFP10 simulations. The noise maps for AliCPT bands are simulated from the noise variance maps. Note that the foreground simulations used in these set of simulations are different from those used in DC1. These simulations are propagated through the GLS pipeline to obtain foreground cleaned $B$-mode maps. These maps are used to obtain the power spectrum and the covariance matrix. In this way we incorporate the variance contribution of the cosmic variance, noise and foreground residuals. In the covariance matrix, we only keep the the elements where the two bins are separated from each other by one or less.

The simulations used to compute the covariance matrix are not TOD filtered as TOD simulation and filtering are computationally expensive. This would imply we would have an additional contribution to the covariance matrix from the filtering. We have verified from CMB and noise only filtered simulations that additional variance due to the filtering and filtering corrections is an subdominant to the leading contributions to the covariance matrix computed here. The filtering effect can only lead to a small correction in the covariance of the first bin. By avoiding the filtering step we can perform 300 simulations which is sufficient for us to compute the covariance matrix with the primary contributions. However, in future analysis we would like to obtain the covariance matrix by propagating sufficient number of filtered maps through the GLS pipeline.

\begin{figure}
    \centering
    \includegraphics[width=0.48\textwidth]{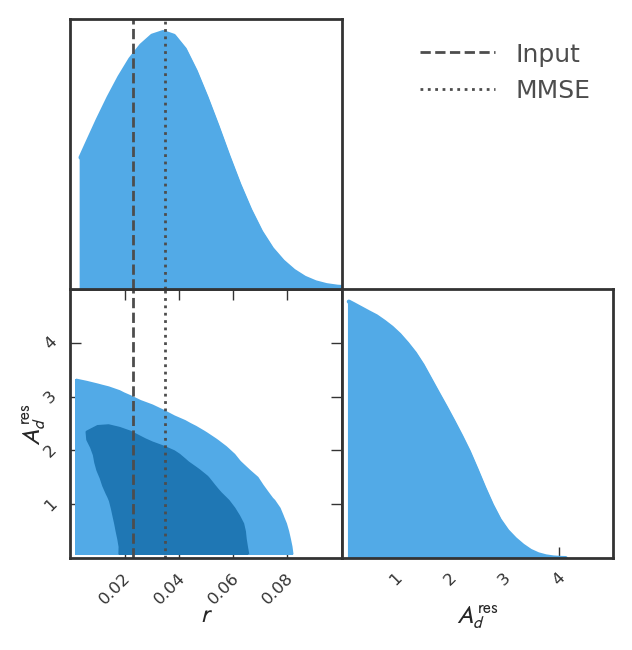}
    \includegraphics[width=0.48\textwidth]{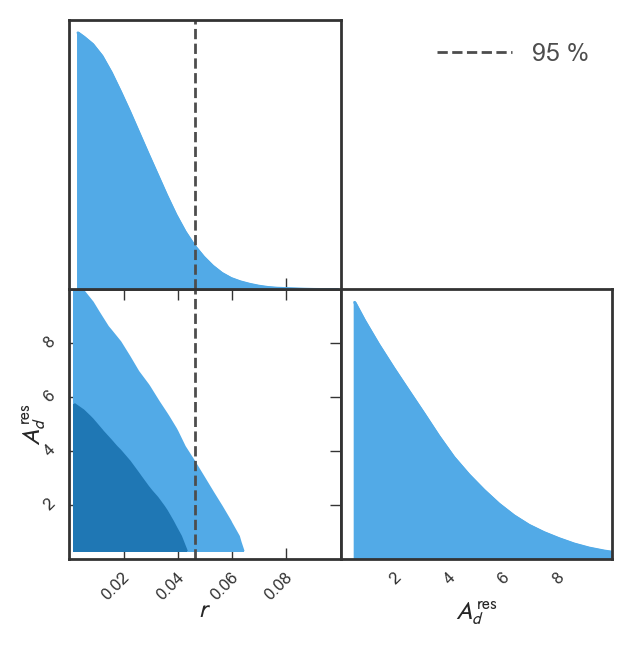}
    \caption{Posterior distribution for the tensor-to-scalar ratio for GLS pipeline in the DC1 case (left) and the null test case (right). The dashed line in the left plot shows the actual DC1 value of $r=0.023$, while the dotted line gives the best-fit value of $r$ which is $r=0.035 \pm 0.020$ for the GLS method. In the right plot, the dashed line indicates the 95\% upper limit of $r<0.043$.}
    \label{fig:GLS_likelihood}
\end{figure}

We fit the $B$-mode power spectrum from the GLS cleaned $B$-map in the multipole range 40-200. For the model used here we adopt an uniform prior on $r$ in the range $0 < r < 1$ and uniform prior on nuisance parameter $A_d^{\rm res}$ in the range $0<A_d^{\rm res}<10$. We use \texttt{emcee} \citep{emcee:2013} to estimate the posterior distribution of $r$. The posterior distribution for $r$ for the GLS pipeline for DC1 case is shown on the left in Fig. \ref{fig:GLS_likelihood}, while the null test is shown on the right of Fig. \ref{fig:GLS_likelihood}. The maximum a posteriori (MAP) best-fit value of $r=0.023^{+0.026}_{-0.013}$ while the minimum mean squared error (MMSE) estimate of $r=0.030_{-0.020}^{+0.019}$ for the GLS pipeline in the DC1 case. In the null test case the posterior distribution maximizes at $r=0$, the 95\% upper limit on $r$ in this case is $<0.043$.

\subsection{cILC pipeline}
\label{ssec:cilc}

Internal linear combination (ILC) is a time tested component separation method that has been used since COBE to obtain foreground cleaned CMB maps from multi-frequency observations \citep{Bennett1992, Bennett:2003, Tegmark:2003, Eriksen2004a, Saha:2006, Saha2008, Delabrouille:2009, Basak:2012, Basak:2013}. The ILC method calculates weights for each frequency channel by minimizing the total variance. The only assumption in the ILC method is the frequency dependence of the CMB (mixing vector), which is unity in all frequency bands if the maps are in thermodynamic temperature units. The traditional ILC methods are blind foreground cleaning methods. They make no assumptions about the frequency scaling of the foreground. If the mixing vector for the CMB is $A_{\nu, \text{ CMB}}$, the traditional ILC weights $W_\nu^{\rm ILC}$ only requires the constraint $\sum_\nu W_\nu^{\rm ILC} A_{\nu, \text{ CMB}} = 1$, to preserve the CMB signal. The ILC weights minimize the contribution of \emph{both} the foreground and noise to reduce the total variance. However, the variance due to foreground and noise in the $B$-mode case for DC1 is too large for traditional ILC methods to minimize them both, leading to large residual foreground signal.

The constrained ILC (cILC) method \citep{Remazeilles2010} is a semi-blind foreground cleaning technique. It uses priors on foreground information to set additional constraints that remove the average dust and synchrotron emission captured by the modeling of the data. For cILC we model the observed data as in Eq. (\ref{eq:model}), where we include models for the average thermal dust and synchrotron emissions. Then we place additional constraints: $\sum_\nu W_\nu^{\rm cILC} A_{\nu, \text{ dust/sync}} = 0$, to remove the average dust and synchrotron emissions from the map. The cILC computes weights by minimizing the variance of the cleaned maps with total three constraints. The constraints on the cILC weights, $W^{\rm cILC}_\nu$ can be written together as:
\begin{equation}
    \sum_\nu W^{\rm cILC}_\nu A_{\nu,c} = e_c,
\end{equation}
where $A_{\nu, c}$ is the same mixing matrix from Eq. (\ref{eq:model}) and $e_c$ is 1 for CMB like for usual ILC method and 0 for the two foreground components modeled. This ensures that for very noisy data, the average dust and synchrotron captured by the model is removed with priority. The cILC weights obtained with these constraints are given by \citep{Remazeilles2010, Remazeilles2021}:
\begin{equation}
    \boldsymbol{W}^{\rm cILC} = \boldsymbol{e} \left( \boldsymbol{A}^t \boldsymbol{C}^{-1} \boldsymbol{A}\right)^{-1}\boldsymbol{A}^t \boldsymbol{C}^{-1},
\end{equation}
with $\boldsymbol C$ as the covariance matrix of the data. The cILC cleaned CMB map is then given by:
\begin{equation}
    \hat s_{\rm CMB} = \sum_\nu W^{\rm cILC}_\nu d_\nu.
\end{equation}

We pre-process and mask the data identically to that for the GLS pipeline discussed in Sec. \ref{ssec:gls}. The cILC is implemented in harmonic space. The detailed implementation of the cILC for the AliCPT DC1 is discussed in Paper III. We obtain $B$-mode power spectrum from the cILC cleaned map. We estimate the noise bias from 50 filtered noise simulations by projecting the noise with the cILC weights. The leakage bias is computed from the cross-spectrum between the Planck-WMAP leakage template and the cILC cleaned $B$-map as discussed in Sec. \ref{sec:filter}. The power spectrum is debiased from the noise and TOD filtering leakage bias. Finally we correct the power suppression with transfer functions. The final power spectrum for cILC is shown in Fig. \ref{fig:GLS-cILC_PS} on the right. 

Similar to the case of GLS pipeline in Eq. (\ref{eq5.8}), we maximize the Gaussian likelihood function of Eq. \eqref{eq:gauss_like} with the theoretical model given by:
\begin{equation}
    C_\ell^{\rm th}(r) = \frac{r}{0.01} C^{\text{tens, }r=0.01}_\ell + A_{\rm lens} C^{\rm lensing}_\ell + A_d^{\rm res} C_\ell^{\rm dust},
\end{equation}
again with $A_{\rm lens}=0$. 
We fit the power spectra in the cILC case multipole range 40-200. The covariance matrix is computed for fiducial power spectrum with $r=0.03$ using 300 simulations prepared in the same way as described in Sec. \ref{ssec:gls}. These maps are propagated through the cILC pipeline to obtain foreground cleaned $B$-mode maps. We then calculate the power spectra and their covariance matrix. As with the GLS pipeline, only elements for bins separated by one or less is retained in the covariance matrix for the likelihood estimation. We obtain the posterior distribution for $r$ by performing an MCMC in the parameter space using \texttt{emcee}. The normalized posterior distribution of $r$ is shown in Fig. \ref{fig:combined_posterior_plot} where the DC1 case is plotted on the left and the null test case is plotted on the right. The MAP best-fit value of $r=0.024^{+0.017}_{-0.015}$ while the MMSE estimate for $r=0.025 \pm 0.016$ in the DC1 case. The posterior distribution for the null test case peaks at $r=0$, and the 95\% upper limit is $<0.050$.



\begin{figure}
    \centering
    \includegraphics[width=\textwidth]{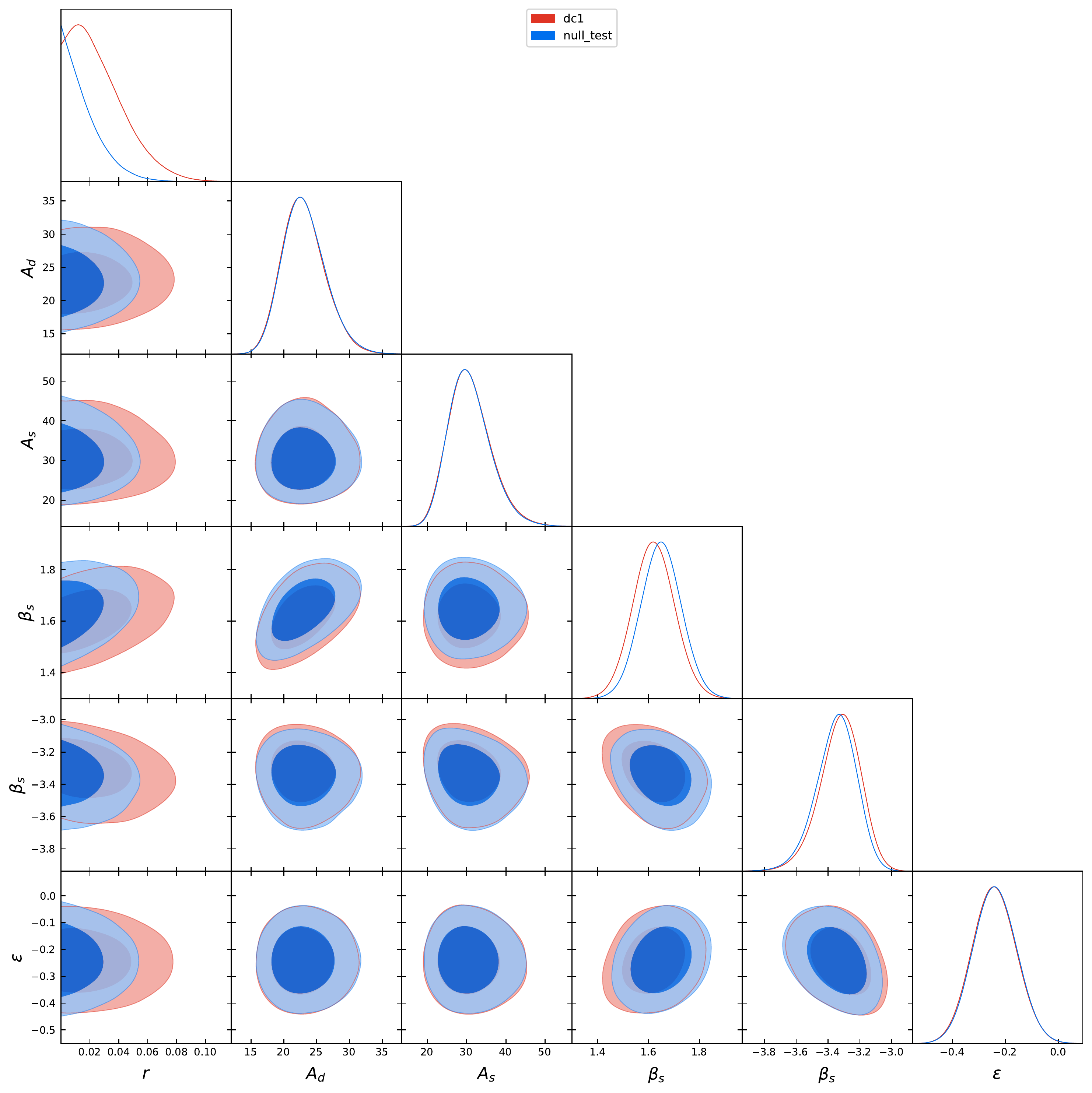}
        \caption{Posterior distribution of a selected subset of parameters of the McMfL pipeline for AliCPT DC1 data (red) and null test case(blue). }
    \label{mcmfl-2d}
\end{figure}

\subsection{McMfL pipeline}
\label{ssec:mcmfl}

In the previous subsections, we perform foreground removal with multi-frequency observations to get the cleaned map (or directly the spectra in case of the ABS), then we estimate the $BB$ power spectra and finally fit the spectra to give the constraint on tensor scalar ratio ($r$). We can also start with our filtering corrected, multi-frequency, auto and cross power spectra and parameterize the foreground emission at the power spectra level, and then constrain the foreground and cosmological parameters simultaneously. This method is conceptually close to the multi-detector multi-component SMICA method \citep{2003MNRAS.346.1089D,2008ISTSP...2..735C} widely used for CMB power spectrum estimation and component separation on various data sets \citep{2005MNRAS.364.1185P,2005A&A...436..785T,2014A&A...571A..12P}, but explicitly uses parametric models of \emph{both} the CMB and the foreground emissions. It has been used for BICEP/Keck and Planck joint analysis \cite{Joint_BICEP&Planck22015}. We implemented a multi-component multi-frequency spectra based likelihood (McMfL) with the parameterized angular power spectra of CMB, uncorrelated galactic dust, uncorrelated galactic synchrotron, and spatial correlation between dust and synchrotron. The total $BB$ auto- and cross-spectra between maps at frequencies $\nu_1$ and $\nu_2$ can be modeled as:
\begin{equation}\label{equ:fg-aps}
\begin{aligned}
    D_{\ell,\nu_1\times \nu_2}^{BB} =&  A_d\Delta_d^{\prime}f_d^{\nu_1}f_d^{\nu_2}\left(\frac{\ell}{80}\right)^{\alpha_d} + A_s\Delta_s^{\prime}f_s^{\nu_1}f_s^{\nu_2}\left(\frac{\ell}{80}\right)^{\alpha_s} + \\ &\epsilon \sqrt{A_d A_s}(f_d^{\nu_1}f_s^{\nu_2}+f_s^{\nu_1}f_d^{\nu_2})\left(\frac{\ell}{80}\right)^{(\alpha_d+\alpha_s)/2} + D_{\ell,{\rm CMB}}^{BB}(r),
\end{aligned}
\end{equation}
where $A_{d}$ ($A_{s}$) is the power amplitude for dust (synchrotron) at pivot frequency ($\nu_{\rm pivot}$) at 353\,GHz (23\,GHz) and pivot angular scale $\ell_*=80$. Here, $\Delta_d^{\prime}$ ($\Delta_s^{\prime}$) describes the decorrelation of the dust (synchrotron) pattern, which we set to 1. The spectral index in $\ell$ domain for dust (synchrotron), $\alpha_d$ ($\alpha_s$) can be written as,
\begin{equation}
    \alpha_{d/s} = \alpha_{d/s, 0} + \alpha_{d/s, r} \ln(\ell/80),
\end{equation}
where $\alpha_{d/s, 0}$ is the scale independent dust/synchrotron spectral index, and $\alpha_{d/s, r}$ is the corresponding running index. The level of spatial correlation between dust and synchrotron is given by $\epsilon$. The factors $f_d^{\nu}$ (resp. $f_s^{\nu}$) scales the dust (resp. synchrotron) power from pivot frequency 353\,GHz (23\,GHz) to observed frequency band with bandpass function, $G(\nu)$, and are defined as follows, 
\begin{align}
    f_d^\nu =& 
    S \times 
    \frac{\int d\nu G(\nu) \nu^{3+\beta_d}\left(\exp(\frac{h\nu}{kT_d})-1\right)^{-1}}
    {\nu_{\rm pivot}^{3+\beta_d}\left(\exp(\frac{h\nu_{\rm pivot}}{kT_d})-1\right)^{-1}} , \label{equ:fdnu} \\
    f_s^\nu =&
    S \times
    \frac{\int d\nu G(\nu) \nu^{2+\beta_s}}
    {\nu_{\rm pivot}^{2+\beta_s}} , \label{equ:fsnu}
\end{align}
with
\begin{equation}
    S \equiv 
    \frac{\nu_{\rm pivot}^4\exp(\frac{h\nu_{\rm pivot}}{kT_{\rm CMB}})\left(\exp(\frac{h\nu_{\rm pivot}}{kT_{\rm CMB}})-1\right)^{-2}}
    {\int d\nu G(\nu)\nu^4\exp(\frac{h\nu}{kT_{\rm CMB}})\left(\exp(\frac{h\nu}{kT_{\rm CMB}})-1\right)^{-2}}. \label{equ:aps_S}
\end{equation}
Here, $\beta_d$ ($\beta_s$) is the dust (synchrotron) spectral index, $T_d$ is the dust temperature, which we set 19.6 K. For the CMB spectra $D_{\ell,{\rm CMB}}^{BB}$, we use \texttt{camb} \citep{Lewis:1999camb} to generate the angular power spectra with tensor-to-scalar ratio ($r$). The parameter space for the likelihood involves \{$r$, $A_d$, $A_s$, $\beta_d$, $\beta_s$, $\alpha_{d,0}$, $\alpha_{d,0}$, $\alpha_{d,r}$, $\alpha_{s,r}$, $\epsilon$\}, while the other cosmological parameters fixed to the best-fit values from Planck 2018 \cite{Planck:2018nkj}. 




We first calculate the bandpowers, ${D}^{BB, \text{ filt}}_{b,\nu_1\times  \nu_2}$, using the PCL-SZ for each $\ell$-bin, for each frequency pair $\nu_1\times \nu_2$ from the observed maps. We set a bin width of $\Delta \ell=25$, and the bin window function is chosen as top-hat function. The BPSapo mask shown in Fig. \ref{fig:apo_mask} is used for all computations. These filtered bandpowers are then corrected for filtering by Eq. \eqref{eq:filter_corr}. 

\begingroup
\renewcommand{\arraystretch}{1.5}
\begin{table}[ht]
    \centering
\begin{tabular} { l  |  r    r  }

\hline 
 Parameter &  DC1  & Null test\\
\hline
{\boldmath$r              $} & $ 0.026\pm 0.019           $ & $< 0.0192                  $\\

{\boldmath$A_d            $} & $23^{+3}_{-4}              $ & $23^{+3}_{-4}              $\\

{\boldmath$A_s            $} & $31^{+4}_{-6}              $ & $31^{+4}_{-6}              $\\

{\boldmath$\alpha_{d,0}   $} & $-0.42\pm 0.58             $ & $-0.41\pm 0.57             $\\

{\boldmath$\beta_s        $} & $1.621\pm 0.083            $ & $1.649\pm 0.080            $\\

{\boldmath$\alpha_{s,0}   $} & $0.56^{+0.81}_{-0.96}      $ & $0.55^{+0.81}_{-0.95}      $\\

{\boldmath$\beta_s        $} & $-3.33^{+0.14}_{-0.11}     $ & $-3.35^{+0.14}_{-0.11}     $\\

{\boldmath$\epsilon       $} & $-0.240\pm 0.083           $ & $-0.239\pm 0.082           $\\

{\boldmath$\alpha_{d,r}   $} & $-2.34^{+1.10}_{-0.86}      $ & $-2.30^{+1.10}_{-0.83}      $\\

{\boldmath$\alpha_{s,r}   $} & $-2.5^{+2.2}_{-1.3}        $ & $-2.4^{+2.2}_{-1.3}        $\\
 
\hline
\end{tabular}

\caption{Constraints and 68\% uncertainties for all model parameters from the McMfL.} 
\label{tab:params_1d_mcmfl}
\end{table}
\endgroup

We adopt the Hamimeche \& Lewis (HL) likelihood approximation \cite{HL} for the McMfL pipeline. The HL likelihood choice is justified as the degrees of freedom per bandpower bin is small due to the filtering and the small sky coverage. The HL likelihood is summarized in Sec. \ref{app:likelihood}, and the likelihood is given in Eq. \eqref{eq:HL_like}. The HL likelihood is insensitive to the chosen fiducial model. The bandpass covariance matrix (BPCM), $\boldsymbol M_{\rm fid}$ only depends on the fiducial model, so we can pre-compute the BPCM. We use the 50 filtered CMB maps generated from our fiducial model ($\Lambda$CDM with tensor ($r=0.01$), which means the fiducial model is free from any foreground emissions) from the Ancillary Data to compute the fiducial power spectra, and the noise simulations in the Ancillary Data to compute the noise only power spectrum.
These fiducial and noise bandpowers are then used to construct the BPCM following a semi-analytic method \cite{re:buza:phd}. We also set the covariance between bandpowers separated by more than one $\ell$-bin to zero. 
The posterior distribution of model parameters is sampled using the modified \texttt{CosmoMC}\footnote{\url{https://cosmologist.info/cosmomc/}}  \citep{Lewis:2002cosmomc}. Uniform prior is adopted for all parameters. The marginalized 2-D posterior distribution of $r$ as well as other foreground parameters is shown in Fig. \ref{mcmfl-2d} and Table \ref{tab:params_1d_mcmfl}. We obtained a best-fit value of $r = 0.026 \pm 0.019$. For the null test case, the constraint on $r$ is $r<0.042$ at $95\%$ confidence level.

\subsection{TF pipeline}
The template fitting (TF) pipeline is an alternate pipeline to perform joint parametric fit of foreground, and filtering parameters along with the tensor-to-scalar ratio, similar to the McMfL pipeline.
The TF and McMfL pipelines share the parametric philosophy of modeling the foreground and the CMB, and the availability of both provides an option to cross-check results obtained with the two pipelines. Despite their similarities they do have some key differences: 1) multiple variants of the foreground parameterization models are adopted in TF (see Paper III), 2) the TF pipeline uses Gaussian likelihood, whereas McMfL uses the HL likelihood, which leads to completely different estimates of the covariance, and 3) the TF uses the `forward modeling' approach for filtering effects (see Sec. \ref{ssec:forward_model}), and so no corrections are applied to the auto- and cross-spectra from multi-frequency maps during analysis. In the TF pipeline the leakage and suppression effects are already included in the CMB tensor-only template and the estimate of $r$ is derived from directly likelihood fitting the template to the DC1 data. 

\begin{figure}
    \centering
    \includegraphics[width=1.0\textwidth]{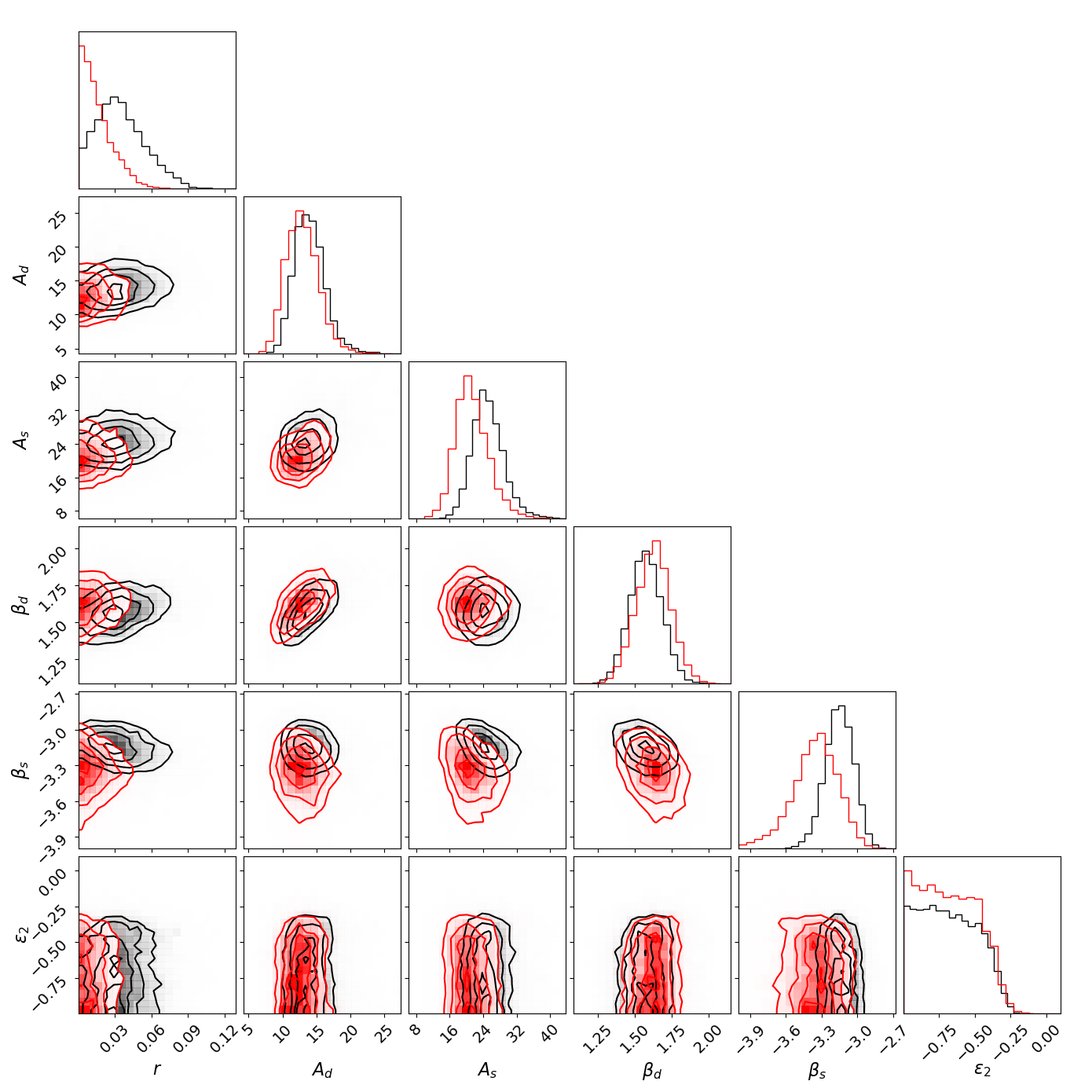}
        \caption{Posterior probability distributions of some selected parameters for the DC1 data (black) and the null test (red), respectively, using the TF pipeline for the foreground model {\it p16} in the range of $20\leq \ell \leq 200$. The off diagonal panels display the joint, marginalized constraint contours at 1.0, 1.5 and 2.0$\sigma$ on parameters, and the diagonal panels display the marginalized distribution of each parameter. }
    \label{fig:p16-r}
\end{figure}

Here we report only the results of the TF pipeline using a more complicated foreground model (dubbed \emph{p16}), as a representative example, different from that of McMfL in Eq. \eqref{equ:fg-aps}. The TF \emph{p16} model including 12 free parameters for the foreground, 3 parameters for the filtering effects on the foreground, together with one CMB parameter $r$ is designed to capture detailed features that we observe from simulations. The added parameters allow us to model the filtering-induced low-$\ell$ suppression for the foreground, the scale-dependence of the synchrotron-dust cross correlation, and the scale-dependence of $\beta_d$ and $\beta_s$. The scale-dependence of the synchrotron-dust cross correlation is modeled by replacing the constant $\epsilon$ in Eq. \eqref{equ:fg-aps} with
$\epsilon(\ell) = \epsilon_2\left({2}/{\ell}\right)^{\alpha_\epsilon}$,
where the uniform priors are adopted on $\epsilon_2\in [-1, 1]$ and $\alpha_\epsilon\in [0, 2]$. Analogously, the scale-dependence of the spectral indices of $\beta_d$ and $\beta_s$ is modeled by replacing constant $\beta_d$ and $\beta_s$ with 
\begin{equation}
  \beta_d \to \beta_d  + \beta_d^\prime \ln({\ell}/{80}), \qquad \beta_s\to \beta_s  + \beta_s^\prime \ln({\ell}/{80}),
\end{equation}
respectively. The filtering-induced suppression factor in the foreground band-power amplitude is modeled as 
\begin{equation}
  F_{\ell} = 1 - f_0\exp\left[-\left({\ell}/{\ell_F}\right)^{\alpha_F}\right]\,,
\end{equation}
with the uniform priors, $f_0\in [0, 1]$, $\alpha_F\in [1, 8]$ and $\ell_F\in [50, 100]$. Here, this suppression factor acts only on the foreground model and not on the CMB, which is modeled through the CMB template defined in Eq. \eqref{eq:cmbt}. Moreover, the filtering-induced suppression factor here is not determined from simulations. Instead we simultaneously estimate the foreground filtering parameters by sampling the total parameter space of the TF posterior. In this respect our approach to the filtering corrections is very different from that of the McMfL pipeline. Due to this difference in foreground modeling, the foreground parameters estimated from the TF, especially with respect to the dust and synchrotron amplitudes, are not guaranteed to be exactly the same as those from the McMfL. 

\begingroup
\setlength{\tabcolsep}{14pt}
\renewcommand{\arraystretch}{1.3}
\begin{table}
\centering
\begin{tabular}{c | r   r}
\hline
Parameter & DC1 & Null test \\
\hline
$\boldsymbol{r}$ & $ 0.035 \pm  0.020 $ & $0.015 \pm  0.012$  \\
$\boldsymbol{A_d}$ & $ 14.15 \pm  2.41$ & $ 12.82 \pm  2.43 $\\
$\boldsymbol{\alpha_d}$ & $ -1.42 \pm  0.49 $ & $ -1.72\pm  0.61 $ \\
$\boldsymbol{\alpha_d^\prime}$ & $ -2.78 \pm  0.80 $& $ -2.59 \pm  1.04 $\\
$\boldsymbol{\beta_d}$ & $ 1.57 \pm  0.10 $& $ 1.62 \pm  0.11 $\\
$\boldsymbol{\beta_d^\prime}$ & $ -0.27 \pm  0.25 $& $ -0.38 \pm  0.30 $\\
$\boldsymbol{A_s}$ & $ 25.15 \pm  3.87 $& $ 21.42 \pm  4.11 $\\
$\boldsymbol{\alpha_s}$ & $ 0.10 \pm  0.54 $& $ -0.24 \pm  0.70 $\\
$\boldsymbol{\alpha_s^\prime}$ & $ -3.08 \pm  0.72 $ & $ -3.23 \pm  0.66 $\\
$\boldsymbol{\beta_s}$ & $ -3.15 \pm  0.12 $& $ -3.36 \pm  0.19 $\\
$\boldsymbol{\beta_s^\prime}$ & $ 0.01 \pm  0.33 $& $ 0.11 \pm  0.47 $\\
$\boldsymbol{f_0}$ & $ 0.57 \pm  0.26 $& $ 0.71 \pm  0.25 $\\
$\boldsymbol{\ell_F}$ & $ 71.34 \pm  11.44 $ & $ 73.69 \pm  11.24 $\\
$\boldsymbol{\alpha_f}$ & $ 4.36 \pm  1.82 $& $ 4.33 \pm  1.67 $\\
$\boldsymbol{\epsilon_2}$ & $ -0.67 \pm  0.18 $ & $ -0.68 \pm  0.19 $\\
$\boldsymbol{\alpha_\epsilon}$ & $ 0.17 \pm  0.08 $ & $ 0.18 \pm  0.10 $\\
\hline
\end{tabular}
\caption{Constraints on the CMB and foreground parameters for {\it p16} model using the TF pipeline, in the cases of the DC1 data and the null test. For each parameter, the mean and the standard deviation of posterior distribution are shown.} 
\label{tab:params_tf}
\end{table}
\endgroup
In the range of $\ell$-bins of interest ($20 \le \ell \le 200$), the TF pipeline uses the auto-/cross-spectral band powers, together with the usual Gaussian likelihood as a good approximation to derive parameter estimates, which is detailed in Eq. \eqref{eq:gauss_like}.  It is important to note that, a non-zero contribution of the foreground-noise covariance between the frequency pairs has to be taken into account (the details of the calculation are given in Paper III), and such contribution will dominate the full covariance at low-$\ell$ bins ($\lesssim60$), leading to an increase in the error bars and giving correct $\chi^2$ values for given degrees of freedom. The apodized mask `ABSapo' is adopted in the TF pipeline, and is discussed in Sec. \ref{app:mask_apo}, and all the band powers are calculated using the PCL-SZ method as discussed in Sec. \ref{sec:Cl_estimators}.



The marginalized estimates on $r$ and foregrounds in the multi-dimensional parameter space are shown in Fig.~\ref{fig:p16-r} for the DC1 data and the null test, where only 5 of 15 foreground parameters are chosen as representatives in order to compare them with the McMfL results clearly. The constraints on all parameters are shown in Table~\ref{tab:params_tf} (the full 2D posterior probability distribution shown in Paper III). As seen, the TF derived $r$ peaks at $r=0.029$, which is slightly greater than that from the McMfL ($r=0.012$) in the DC1 case, although this discrepancy is consistent within the 1-$\sigma$ level of $\sigma(r)\approx0.021$. Additionally, in comparison with the McMfL-derived error on $r$, increasing the free foreground parameters from 9 to 15 does not significantly increase its statistical error, which indicates that the TF pipeline and its parameterization method are robust and effective.  

The measured values of the dust and synchrotron spectral indices, $\beta_d$ and $\beta_s$, from both TF and McMfL pipelines are in a good agreement. In addition, we do observe a negative correlation ($\epsilon_2<-0.037$ at $95\%$CL), which is the same as the McMfL derived value using the single parameter $\epsilon$ as parameterization. We also find that the amplitudes of both synchrotron and dust, $A_s$ and $A_d$, are significantly smaller than those of the McMfL model, which should be due to our different treatments for the filtering effects, where the TF utilizes the smooth function $F_\ell$ to fit the filtering suppression, while the McMfL corrects such suppression directly on the observed band powers. By a further examination, both the TF and McMfL models can reconstruct the band powers of the 23 and 353 GHz data quite well according to their best-fit parameters.

\begin{figure}
    \centering
    \includegraphics[width=0.49\textwidth]{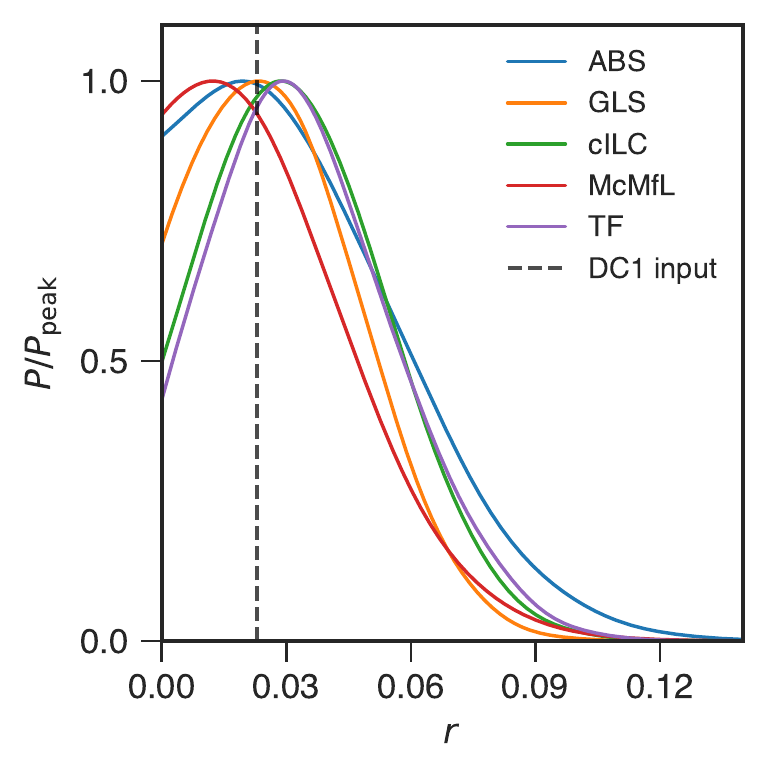}
    \includegraphics[width=0.49\textwidth]{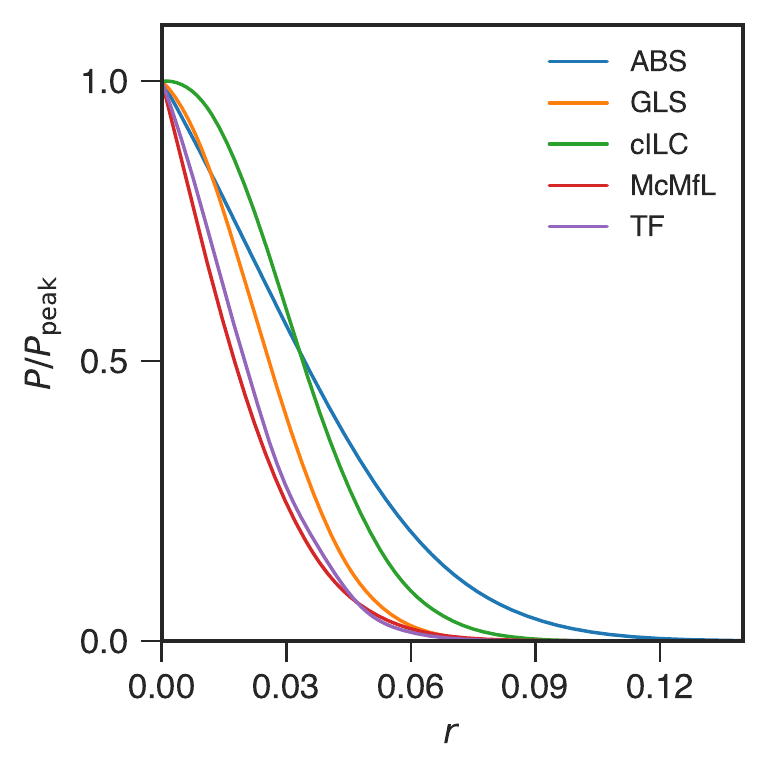}
    \caption{Plots showing the normalized posterior distribution of the tensor-to-scalar ratio ($r$) for ABS (blue), GLS (orange), cILC (green), McMfL (red) and TF(purple) pipelines. In the left plot we show the DC1 case where the true value of $r=0.023$ is shown with the black dashed line and on the right we have plot for the null test case.}
    \label{fig:combined_posterior_plot}
\end{figure}

\section{Discussion and conclusions}
\label{sec:conclusions}
In the previous section we have summarized the working principles and performances of our five data analysis pipelines. A comparison of the normalized marginal posterior distribution for $r$ obtained with the different pipelines is shown in Fig. \ref{fig:combined_posterior_plot}. We find that all our methods have similar posteriors with comparable widths. For single season AliCPT data jointly analysed with Planck HFI and WMAP K band, our sensitivity on $r$, $\sigma(r)$, ranges between $0.019$ and $0.025$. For the DC1 case plot of Fig. \ref{fig:combined_posterior_plot} we find posteriors from all pipelines peak above $0$ and the maximum a-posteriori (MAP) estimate of $r$ ranges between $0.012$ and $0.029$, which include the true $r$. For all pipelines the results for DC1 are consistent with 0 at 2$\sigma$ level. We find the the expectation value of $r$ to be a bit higher than the MAP estimate. This is partly contributed by the asymmetric prior of $[0,1]$. We conclude that our methods do not show any obvious bias as the posterior peak is unbiased.

\begin{table}[b]
    \centering
    \begin{tabular}{l|c c c c c} 
        \hline \\[-1ex]
         \textbf{Method} & ABS & GLS & cILC & McMfL & TF (p16)\\[1ex] 
         \hline \\[-1ex]
         $\boldsymbol{r}$ [MAP] & $0.019$ & $0.023$ & $0.024$ & $0.012$ & $0.029$\\[1.5ex] 
         $\ell~{\rm range}$ 
         & [20,200]
         & [40,200]
         & [40,200]
         & [50,250] 
         & [20,200] \\
          $\chi^2_{\rm min }/{\rm DOF}$
         & 4.3/8
         & 2.4/3
         & 2.9/3
         & 219.6/214
         & 248.3/236\\        
         $\boldsymbol{r}$ [MMSE] & $0.036^{+0.025}_{-0.025}$ & $0.030^{+0.019}_{-0.020}$ & $0.025^{+0.016}_{-0.016}$ & $0.026^{+0.019}_{-0.019}$ & $0.035^{+0.021}_{-0.021}$\\[1ex] 
        \hline
    \end{tabular}
    \caption{Comparison of the tensor-to scalar ratio ($r$) estimate from the different methods for the DC1 input of $0.023$. In the first row we show maximum a posteriori (MAP) estimate of $r$ (where the posterior is maximum), the second and third rows are the multipole range and the chi-squired value / degrees of freedom (DOF) for the best-fit models,
    and in last row we show the minimum mean squared error (MMSE) estimate of $r$ (posterior weighted average of $r$).}
    \label{tab:DC1_r_estimate_comparison}
\end{table}

\begin{table}[ht]
    \centering
    \begin{tabular}{l|c c c c c} 
        \hline \\[-1ex]
         \textbf{Method} & ABS & GLS & cILC & McMfL & TF (p16)\\[1ex] 
         \hline \\[-1ex]
         
         $\boldsymbol{r}$ (95\% CL) & $< 0.069$ & $<0.043$ & $<0.050$ & $<0.042$ & $<0.041$\\[1ex] 
        \hline
    \end{tabular}
    \caption{Comparison of the tensor-to-scalar ratio ($r$) 95\% upper limit from the different methods for the null test with noise and foreground only. For all cases the posterior distributions peak at $r=0$.}
    \label{tab:nulltest_r_95pcCL_comparison}
\end{table}

If we now consider the null test case, where we study biases from noise and foreground, we find the posteriors for the null test case shown in Fig. \ref{fig:combined_posterior_plot}, peak at 0 for all pipelines. As before the spread is comparable for the different pipelines. In Table \ref{tab:nulltest_r_95pcCL_comparison} we show the 95\% upper limit on $r$ in this case. From foreground and noise only case we are able to set upper limits on $r$ between $0.041$ and $0.069$. We note that due the low signal-to-noise, noise debiasing with just 50 simulations is a challenge, which may lead to spurious biases in the results of some pipelines. {We intend to investigate this issue in greater detail with a larger set of simulations, in a future study.}
We still find that our pipelines, despite the variety in their analysis principles, give robust estimates that consistent with each other.

Our current simulations have included some of the systematic effects that we expect to find in the AliCPT data, there are several other systematics like bandpass mismatch, $T$ to $P$ leakage, non-gaussian beams, that we have not included in the present simulations. So we expect that other systematic issues will push the uncertainties higher. We would point out that the leading contribution to $\sigma(r)$ is from the noise and foreground residuals which we have already included in the analysis.

In this work we have demonstrated the performance of the data analysis pipelines for AliCPT-1. The data challenge and null test results presented here are for 4 detector modules and a single season of observation. While the AliCPT project will span multiple years and will be upgraded with more detector modules, our results here should provide a good estimate of the initial sensitivity on $r$ that can be achieved.

\acknowledgments
This work is supported by the National Key R\&D Program of China Grant (No. 2021YFC2203100, 2020YFC2201600), NSFC No. 12273035, 11903030 and 12150610459, the Fundamental Research Funds for the Central Universities under Grant No. WK2030000036 and WK3440000004. Some of the results in this paper have been derived using the HEALPix \citep{Healpix2005} package. Based on observations obtained with Planck (\href{http://www.esa.int/Planck}{http://www.esa.int/Planck}), an ESA science mission with instruments and contributions directly funded by ESA Member States, NASA, and Canada.

\appendix
\section{Masks and apodization}
\label{app:mask_apo}
 {\color{black} Since different pipelines have the quite different treatments on various contamination, we should adopt different masks in the analysis.}
In this section we summarize the mask and mask apodization choices that are adopted in this paper. First, the choice for our binary masks are outlined, followed by point source masks and last we discuss the apodized masks.

\emph{Binary masks:} There are three different binary masks used in this work. The ABS mask has $f_{\rm sky}=5\%$. 
It is produced by masking the sky which has smaller than 20\% of the maximum hit count and also removing higher foreground areas. The `ABS' mask is used in both the ABS and the TF pipelines. The `UNP' mask is produced by removing any pixel with noise standard deviation {\color{black}{larger than}} 20 $\mu$K-pixel at \texttt{NSIDE}=1024 for 150 GHz channel, and removing the sky above declination of $65^\circ$ to remove foregrounds. This mask with $f_{\rm sky}=6.7 \%$, 
and is used in the GLS and cILC pipelines. The `Bmsk' is constructed to keep the $7\%$ deepest sky based on AliCPT 150 GHz channel. This mask is used in the McMfL analysis.

\emph{Point source masks:} The ABS and TF pipelines do not treat the point sources separately. The GLS and cILC pipelines share a common pre-processing stage to mask point sources and inpaint the point source masking. This point source pre-processing stage is discussed in detail in Paper III. The point source mask, `PS1' used in this case is obtained from unfiltered HFI 100 GHz map using the filter defined in \citep{Tegmark:1998}. The McMfL uses existing polarized intensity maps for point sources at 90 GHz to compute the point source mask, `PS2'. All point sources above 3 $\mu$K are masked in PS2. 
\begin{figure}
    \centering
    \includegraphics[width=0.3\textwidth, trim= 0.7cm 0.4cm 0.7cm 0.4cm, clip]{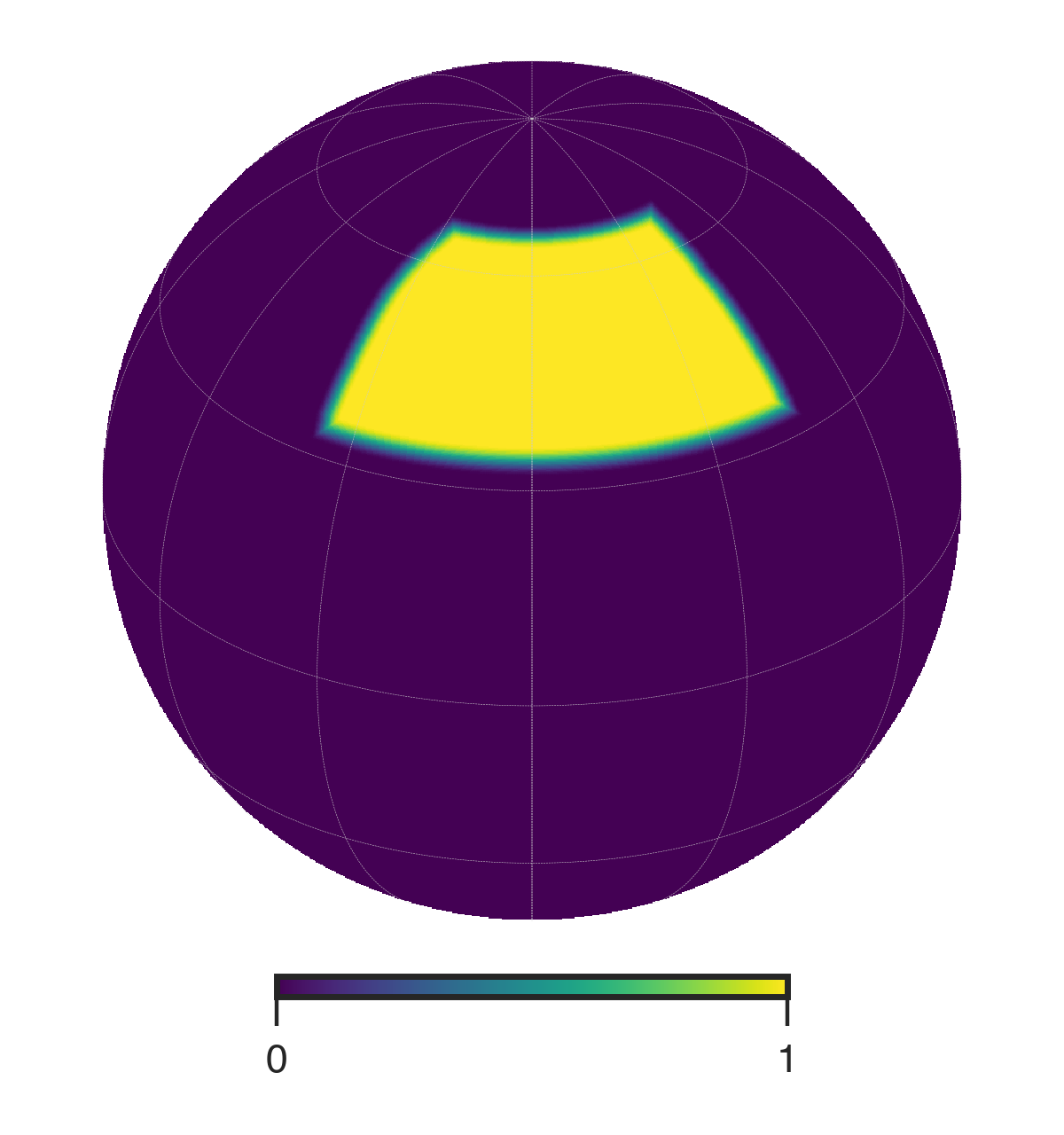}
    \includegraphics[width=0.3\textwidth]{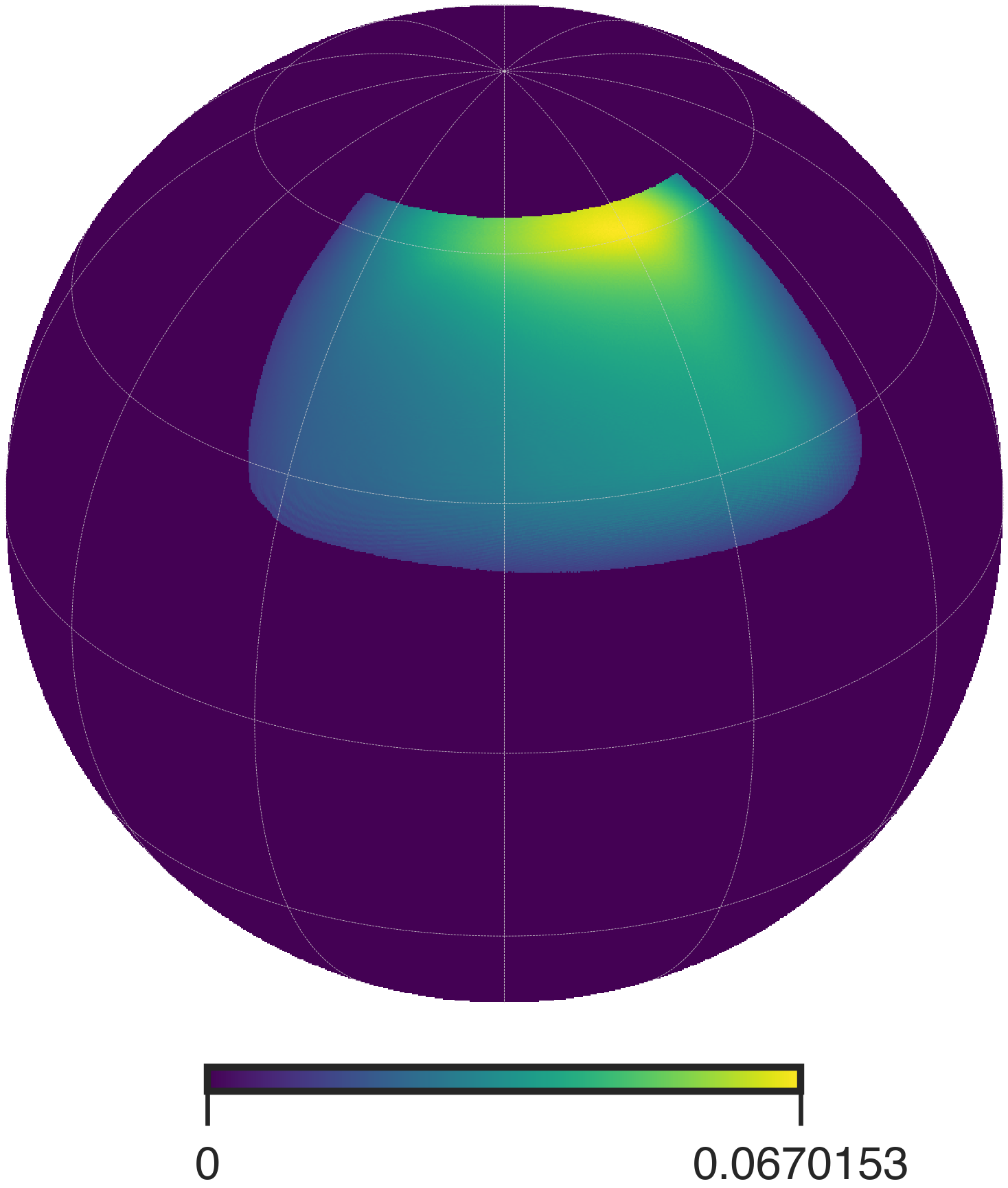}
    \includegraphics[width=0.3\textwidth,, trim= 0.7cm 0.4cm 0.7cm 0.4cm, clip]{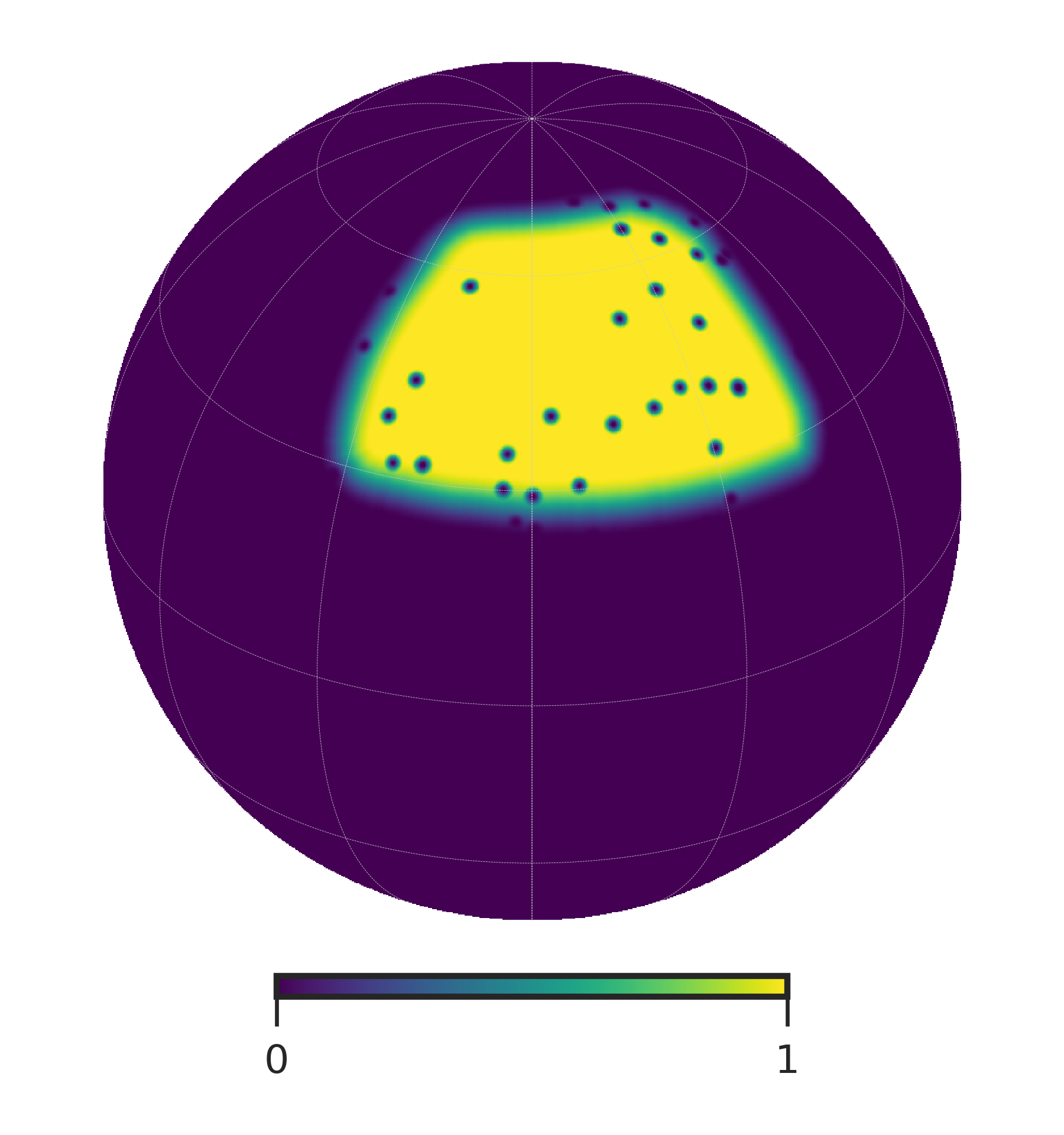}
    \caption{On the left we have the $4^\circ$ \texttt{C2} apodized `ABSapo' mask used in the ABS and TF pipelines, in the middle we have the inverse noise variance weighted `UNPinv' mask used in the GLS and cILC analysis and on the right is the $8^\circ$ and $1.5^\circ$ \texttt{C2} apodized `BPSapo' used in the McMfL pipeline.}
    \label{fig:apo_mask}
\end{figure}

\emph{Apodized masks:} The ABS, McMfL and TF pipelines use the \texttt{C2} apodization to smooth the edges of the mask. Here we adopted the Namaster definition for the \texttt{C2} apodization function, which gives the multiplicative factor, $f$, for a pixel as:
\begin{equation}
f=\left\{\begin{array}{cc}
\frac{1}{2}\left[1 - \cos (\pi x)\right] & x<1 \\
1 & \text { otherwise }\,,
\end{array}\right.
\end{equation}
where $x=\sqrt{(1-\cos \theta) /\left(1-\cos \theta_{*}\right)}$, $\theta$ is the angular separation of the pixel from the nearest masked pixel and $\theta_*$ is the apodization length. The ABS mask is apodized with $\theta_*=4^\circ$, \texttt{C2} apodization. This `ABSapo' mask is used in both the ABS and the TF pipelines. The ABSapo mask is shown on the left in Fig. \ref{fig:apo_mask}. The GLS and cILC method adopts inverse noise variance weighted mask for patch defined by the UNP binary mask. We denote this mask by `UNPinv' and it is shown in the middle plot of Fig. \ref{fig:apo_mask}. For the McMfL pipeline the apodized mask is prepared as follows: 
\begin{enumerate}
    \item We apodize the binary mask `Bmsk' with $\theta_*=8^\circ$, \texttt{C2} apodization. Let us denote this by `apo1'.
    \item The PS2 point source mask is apodized with $\theta_*=1.5^\circ$, \texttt{C2} apodization to produce `apo2'.
    \item Multiplying both these mask we finally produce `BPSapo' which is ${\rm apo1}\times {\rm apo2}$. 
\end{enumerate}
The BPSapo is used in the analysis of the McMfL pipeline and is shown on the right in Fig. \ref{fig:apo_mask}.

\section{Likelihood functions}
\label{app:likelihood}
In this work either the Gaussian likelihood or the Hamimeche \& Lewis (HL) likelihood has been adopted for parameter estimation. We will define them here.

\emph{Gaussian likelihood:} The gaussian likelihood is defined as:
\begin{equation}
    - 2 \ln \mathcal{L}(r, \theta) = \sum_{b b'} \left[\hat X_b - X_b^{\rm th}(r, \theta)\right]\left[M^{-1}_{\rm fid}\right]_{ bb'} \left[\hat X_{b'} - X_{b'}^{\rm th}(r, \theta)\right],
    \label{eq:gauss_like}
\end{equation}
where $b, b'$ are indices for the multipole bins, $\hat X_b$ is either binned $\hat D_\ell^{BB}$ in the ABS and TF pipelines or binned $\hat C_\ell^{BB}$ for the GLS and cILC pipelines, $X_b^{\rm th}(r, \theta)$ denote the binned power spectrum from theory. Here $\theta$ denotes all other model parameters other than the tensor-to-scalar ratio. The fiducial covariance matrix \cite{newref}, $[\boldsymbol M_{\rm fid}]_{bb'}=\langle(X_{{\rm fid,}b} - \langle X_{{\rm fid,}b}\rangle) (X_{{\rm fid,}b'} - \langle X_{{\rm fid,}b'}\rangle)\rangle$. In case of the TF pipeline $\boldsymbol M_{\rm fid}$, is the full covariance matrix of the auto- and cross-spectra between different frequency channels. 

\emph{Hamimeche $\&$ Lewis likelihood:} Due to the effects of TOD filtering, and partial sky, the degrees of freedom per bandpower is small, especially for low $\ell$-bins. For low degrees of freedom the Hamimeche \& Lewis (HL) likelihood approximation \cite{HL} is shown to have advantages over the gaussian likelihood. This is defined as: 
\begin{equation}
    -2\ln\mathcal{L} = \sum_{b,b'}[\boldsymbol X_g]_b^t [\boldsymbol M_{\rm fid}]^{-1}_{b,b'} [\boldsymbol X_g]_{b'},
    \label{eq:HL_like}
\end{equation}
where
\begin{equation}
    [\boldsymbol X_g]_b = {\rm vecp}\left( [\boldsymbol C_{\rm fid}]_b^{1/2}\boldsymbol g \left(\boldsymbol C^{\rm th}_b(r, \theta)^{-1/2}\hat{\boldsymbol C}_b \boldsymbol C^{\rm th}_b(r, \theta)^{-1/2} \right)[\boldsymbol C_{\rm fid}]_b^{1/2}\right).
\end{equation}
Here $\boldsymbol C^{\rm th}_b(r,\theta)$ is the theoretical prediction for the bandpower matrix depending on the model parameters for each $\ell$ bin, $[\boldsymbol C_{\rm fid}]_b$ is fiducial model bandpower matrix, and  $\hat{\boldsymbol{C}}_b$ is the observed bandpower matrix computed from the data, and all the bandpower matrices include the noise contribution. The quantity ${\rm vecp}(\boldsymbol A)$ is a vector of $n(n+1)/2$ distinct elements of matrix $\boldsymbol A$, where $n$ is the dimension of $\boldsymbol A$, and $\boldsymbol g(\boldsymbol A)$ is a matrix operation that applies the function $g(x)={\rm sign}(x-1)\sqrt{2(x-\ln(x)-1)}$ to the eigenvalues of matrix $\boldsymbol A$. The bandpower covariance matrix (BPCM) for the fiducial model is given by:
\begin{equation}
    [\boldsymbol M_{\rm fid}]_{b,b'} = \left<\left([\boldsymbol{\hat X}_{\rm fid}]_b-[\boldsymbol X_{\rm fid}]_b\right)\left([\boldsymbol{\hat X}_{\rm fid}]_{b'}-[\boldsymbol X_{\rm fid}]_{b'}\right)^t\right>.
\end{equation}
\bibliographystyle{jhep}
\bibliography{references}

\end{document}